\documentclass[aps,pre,twocolumn,amsmath,amssymb,floatfix,superscriptaddress,nofootinbib]{revtex4-2}
\usepackage{graphicx}
\usepackage{amsmath}
\usepackage{placeins}
\usepackage{bm}
\usepackage[dvipsnames]{xcolor}
\usepackage{amssymb}
\usepackage{mathptmx}
\usepackage{hyperref}
\usepackage{array}
\usepackage{balance}
\usepackage[normalem]{ulem}

\newcommand*{\br}{{\bf r}}

\newcommand*{\be}{{\bf e}}
\newcommand*{\bn}{{\bf n}}

\newcommand*{\bv}{{\bf v}}
\newcommand*{\bs}{{\bf s}}
\newcommand*{\bV}{{\bf V}}

\newcommand*{\bF}{{\bf F}}

\newcommand*{\bOmega}{{\boldsymbol{\Omega}}}

\newcommand{\eq}[1]{Eq.~(\ref{#1})}

\newcommand{\fig}[1]{Fig.~\ref{#1}}

\newcommand{\fref}[1]{Fig.~\ref{#1}}

\begin{document}

\title{Chemotactic behavior for a self-phoretic Janus particle near a patch source of fuel}

\author{Viviana Mancuso}
\affiliation{Department of Mechanical Engineering, University of Hawai’i at 
M{\=a}noa, 2540 Dole Street, Holmes Hall 302, Honolulu, Hawaii 96822, USA}
\affiliation{International Institute for Sustainability with Knotted Chiral Meta Matter (WPI-SKCM2),
Hiroshima University, Higashi-Hiroshima, Hiroshima 739-8526, Japan}

\author{Mihail N. Popescu}
\email{mpopescu@us.es}
\affiliation{Department of Atomic, Molecular, and Nuclear Physics, University of Seville,
41080 Seville, Spain}

\author{William E. Uspal}
\email{uspal@hawaii.edu}
\affiliation{Department of Mechanical Engineering, University of Hawai’i at 
M{\=a}noa, 2540 Dole Street, Holmes Hall 302, Honolulu, Hawaii 96822, USA}
\affiliation{International Institute for Sustainability with Knotted Chiral Meta Matter (WPI-SKCM2),
Hiroshima University, Higashi-Hiroshima, Hiroshima 739-8526, Japan}

\begin{abstract}
Many biological microswimmers are capable of chemotaxis, i.e., they can sense an ambient chemical gradient and adjust their mechanism of motility to move towards or away from the source of the gradient. Synthetic active colloids endowed with chemotactic behavior hold considerable promise for targeted drug delivery and the realization of programmable and reconfigurable materials. Here we study the chemotactic  behavior of a Janus particle, which converts ``fuel'' molecules, released at an axisymmetric chemical patch located on a planar wall, into ``product'' molecules at its catalytic cap and moves by self-phoresis induced by the product. The chemotactic behavior is characterized as a function of the interplay between the rates of release (at the patch) and the consumption (at the particle) of fuel, as well as of details of the phoretic response of the particle (i.e., its phoretic mobility). Among others, we find that, under certain conditions, the particle is attracted to a stable ``hovering state'' in which it aligns its axis normal to the wall and rests (positions itself) at an activity-dependent distance above the center of the patch. 
\end{abstract}

\date{\today}
\maketitle

\section{Introduction}
Synthetic microswimmers have gained significant attention in recent years due to their potential applications in various fields \cite{ebbens2016active}, including materials science \cite{aubret2021metamachines}, micro/nanotechnology \cite{baraban2012transport}, environmental remediation \cite{wang2019photocatalytic,hermanova2022micromachines}, and biomedicine \cite{gao2014synthetic,chen2024enzymatic}. Among synthetic microswimmers, catalytically active Janus particles (JPs), capable of autonomous motion in response to self-generated chemical gradients, present a particularly intriguing avenue for exploration \cite{howse2007self}. These particles consume molecular ``fuel'' (i.e., reactant) available in the surrounding liquid solution by catalyzing a chemical reaction over a region of their surface. The free energy of the chemical reaction is used to induce the particle's mechanical motion through an interfacial molecular mechanism known as phoresis \cite{derjaguin,anderson1989colloid,moran2017phoretic}. In brief, gradients in the chemical composition of the solution along the surface of a particle, in conjunction with molecular-scale interactions between the particle surface and the various chemical species (solvent, reactant, and reaction product) present in the solution, drive the hydrodynamic flow of the solution, leading to directed ``swimming'' motion of the particle \cite{derjaguin,anderson1989colloid}.

For catalytic Janus particles,  the reaction rate, and therefore the interfacial swimming actuation, depends on the local concentration of  molecular fuel \cite{howse2007self}. This observation raises intriguing possibilities for controlling particle behavior. In particular, an understanding of how these particles  respond to external gradients of fuel concentration may allow for realization of artificial chemotaxis, inspired by chemotaxis in biological micro-organisms. Here, one can make a suggestive analogy: in order to move towards regions with a higher concentration of nutrients or other chemoattractants, micro-organisms sense local concentrations and adjust their locomotion mechanisms accordingly.  A classical example is \textit{E. coli}, which can migrate towards a food source in liquid solution by temporally sampling the local nutrient concentration, and, acting on these measurements, suitably modulating a random sequence of straight line ``runs'' and uncontrolled ``tumbling'' events that randomize the bacterium's swimming direction \cite{berg2004coli}. While \textit{E. coli} performs temporal sampling, eukaryotic cells are large enough to directly sense chemical gradients across their body length using surface receptors \cite{berg1977physics,levine2013physics,insall2022steering}. For instance, for the amoeba \textit{Dictyostelium}, an asymmetric distribution of bound and unbound surface receptors can trigger polarization of internal biochemical pathways, allowing this organism to steer towards a food source \cite{levine2013physics}. Neutrophils can chemotax towards an inflammation site through similar mechanisms \cite{iglesias2008navigating,petri2018neutrophil}. Although \textit{Dictyostelium} and neutrophil motility usually involves contact with a solid surface, chemotaxis has also been observed for these cells when freely suspended and swimming in solution \cite{barry2010dictyostelium}.

Motivated by the swimming-squirming analogy between the motility of catalytic Janus particles and that of micro-organisms \cite{golestanian2007designing,Eskandari2018,maity2022unsteady}, many studies have considered whether such analogy extends towards an equivalent of chemotaxis of synthetic swimmers exposed to spatial gradients in the concentration of reactant. Here, we highlight a few of these studies; comprehensive reviews are provided by Refs. \citenum{huang2021designing} and \citenum{xia2024chemotactic}. In an early work, Hong \textit{et al}. studied the behavior of Pt/Au bimetallic nano-rods in the vicinity of a gel soaked with hydrogen peroxide ``fuel.''  The rods were observed to accumulate at the gel \cite{hong2007chemotaxis}. Subsequently, Byun \textit{et al}. pointed out the possibility of hydrodynamic flows sourced by the fuel patch in this set-up, and developed a framework for distinguishing advective and self-propulsive contributions to particle motion \cite{byun2017distinguishing}. Baraban \textit{et al}. studied chemotactic behavior of tubular micro-jets and Pt@SiO2 Janus spheres in a microfluidic device\cite{baraban2013chemotactic}. A solution containing hydrogen peroxide  was continuously injected in one inlet port of the device, leading to a gradient of chemical ``fuel'' perpendicular to the direction of flow. The microswimmers exhibited a tendency to orient towards the region of high fuel concentration and accumulate in that region. In a similar microfluidic set-up, catalase-coated and urease-coated liposomes were observed to migrate towards and away from, respectively, regions of high substrate concentration \cite{somasundar2019positive}. More recently, Xiao \textit{et al}. used stop-flow microfluidic gradient generation to study chemotactic behavior of Cu@SiO2 Janus spheres, finding a tendency of these particles to align their cap towards higher fuel concentration \cite{xiao2022platform}. By stopping the flow, they could eliminate any potential contribution of surface-assisted cross-stream migration to particle motion \cite{katuri2018cross}. In an effort to realize chemotactic active colloids that use biocompatible fuels, Mou \textit{et al}. fabricated ZnO-based Janus spheres fuelled by dissolved carbon dioxide \cite{mou2021zno}. They observed these Janus particles to chemotax towards a carbon dioxide source through reorientation. Additionally, Zhou \textit{et al.} studied bottle-shaped particles powered by internalized enzymatic decomposition of glucose \cite{zhou2022torque}. Similarly, they found that ``opening'' of a bottle will rotate towards a glucose source, leading to chemotactic migration. 

To shed some light on mechanisms underlying experimental results, Popescu \textit{et al.} presented a framework for understanding how the microscopic coupling between the orientation of a self-phoretic Janus particle and an ambient chemical gradient can lead to chemotaxis, defined to be active reorientation of the swimming direction with respect to the  gradient \cite{popescu2018chemotaxis}. Active reorientation distinguishes chemotaxis from chemokinesis, defined as a variation of particle speed due to ambient chemical gradients \cite{popescu2018chemotaxis,moran2021chemokinesis}. As a microscopic mechanism that induces reorientation, Popescu \textit{et al.} pinpointed a crucial role for a quantitative difference in chemi-osmotic response of the catalytic and inert ``faces'' of the particle to a chemical gradient\cite{popescu2018chemotaxis}, which is the equivalent for the active particles of the classic result of Anderson concerning the electrophoresis of colloids with non-uniform zeta potential \cite{anderson1989colloid}. The importance of a contrast in phoretic mobilities of the two ``faces'' for inducing alignment of the particle with an externally maintained gradient had also been noted in the context of thermophoretic Janus particles \cite{bickel2014polarization}.  Building upon the analysis by Saha \textit{et al.}\cite{saha2014clusters}, T{\u{a}}tulea-Codrean and Lauga presented a comprehensive theoretical model for the motion of a self-phoretic Janus particle in a linear background reactant gradient in unconfined solution \cite{tuatulea2018artificial}. They assumed that the reaction rate is proportional to the reactant concentration (i.e., first-order kinetics), considered the phoretic response to both the reactant and product gradients, and developed comprehensive analytical expressions for the translational and angular velocity of a Janus particle with arbitrary orientation with respect to the background gradient. Using these results, they considered the dispersion of a dilute suspension of such Janus particles exposed to a linear gradient in fuel. Finally, we mention that chemotaxis of chemically-powered micromotors in a fuel gradient has also been studied by means of mesoscopic particle-based simulations \cite{chen2016chemotactic,deprez2017passive}.

Moving from the single-particle behavior to the context of suspensions, chemotaxis has been noted to play a role in phase separation, clustering, and other collective phenomena in systems of self-phoretic active colloids \cite{stark2018artificial}.  When these particles interact through self-generated chemical gradients, the sign of chemotactic alignment, i.e., the attractive or repulsive character, determines whether these systems form finite-size clusters, or collapse into a single large cluster \cite{pohl2014dynamic}. Saha \textit{et al}. distinguished four modes of response of an active colloid to an ambient chemical gradient, including chemotactic alignment, and mapped out a phase diagram of collective behaviors induced by chemotaxis \cite{saha2014clusters}. Chemotactic alignment can cooperate or compete with other interparticle interactions, especially activity-sourced hydrodynamic interactions \cite{lushi2012collective,huang2017chemotactic,traverso2020hydrochemical}. For instance, it has been shown that activity-induced fluid stirring can disrupt chemotactic pattern formation \cite{traverso2020hydrochemical}.

Preceding theoretical studies have laid the groundwork for understanding  chemotaxis of catalytic Janus particles. However, important aspects of the problem still remain to be explored. For instance, most experiments involve a spatially localized and finite-sized source of fuel (i.e., a ``patch.'') A patch will generate a fuel concentration field that has nonlinear dependence on spatial position. It is only far away from the patch, i.e., for distances from the patch that are much larger than the characteristic length scale of the patch, that a linearly varying field is a good approximation. However, nonlinear gradients may play a significant role in determining particle behavior, especially as the Janus particle approaches the patch. Secondly, the interplay of the characteristic sizes of the patch and particle may be important, especially as the particle approaches the patch. Thirdly, in experimental realizations, the patch is usually associated with a confining boundary,  such as when the patch is the surface of another colloidal particle, or when the patch is imprinted on a planar substrate. A solid confining boundary will modify the chemical and hydrodynamic fields sourced by patch and the Janus particle, affecting particle motion. Here, we recall that even an inert planar wall can induce wall-bounded ``sliding'' and ``hovering'' states of a catalytic Janus particle \cite{uspal2015self}.

In this study, we investigate the motion of a Janus particle activated by a finite-sized, spatially localized patch source of ``fuel'' in a confining geometry. Specifically, we consider the motion of a catalytic Janus particle near a chemical patch with axisymmetric shape that continuously emits chemical fuel. The model assumes first-order chemical kinetics and therefore resolves the dependence of the catalytic activity of the particle on the local concentration of reactant and motion by self-phoresis induced \textit{solely} by the product. Using first-order kinetics, the model can probe the relative significance of reaction rate (rate of fuel consumption by the particle) and fuel diffusion \cite{cordova2008osmotic,michelin2014phoretic}. Moreover the model also explicitly considers the sizes of the particle and the patch, thus going beyond a far-field point particle and point source analysis (which is recovered in the limiting case of large distances). 
 In large part, this study will focus on an axisymmetric configuration of the particle and patch, in which the particle center is located above the patch center, and the particle axis is aligned with the patch normal, with the general aim of understanding the conditions under which a Janus particle will navigate towards or away from the patch. Although this focus will limit our direct consideration of chemotactic reorientation, it is the logical starting point for understanding general three-dimensional motion.

In summary, our study provides foundations for understanding how a Janus particle responds to a localized, not point-like, source of fuel. The distinctive features include the consideration of a non-uniform reactant concentration field within the solution, sourced by a chemically active patch, as well as the role of confining geometry. Our main finding is that, under certain conditions, these features induce a novel ``hovering'' state in which the particle remains motionless at a steady height above the patch on the wall. The precise value of the hovering height depends on the dimensionless reaction rate (Damköhler number), as well as the material properties of the particle (phoretic surface mobility) and the details of the geometry (the shape of the particle, the shape of the patch, the size ratio, etc.) Depending on these parameters, the hovering state can be stable or unstable against vertical perturbations of the particle position. We also briefly investigate the stability of the hovering state against general three-dimensional 3D motions. We show that the hovering state can indeed be stable in 3D, and attract a particle from an arbitrary initial position and orientation that is not an axisymmetric configuration. Therefore, our study provides proof of concept that a fuel-releasing chemical patch can attract a Janus particle through chemotaxis. 

\begin{figure}[h]
\centering\includegraphics[width=\linewidth]{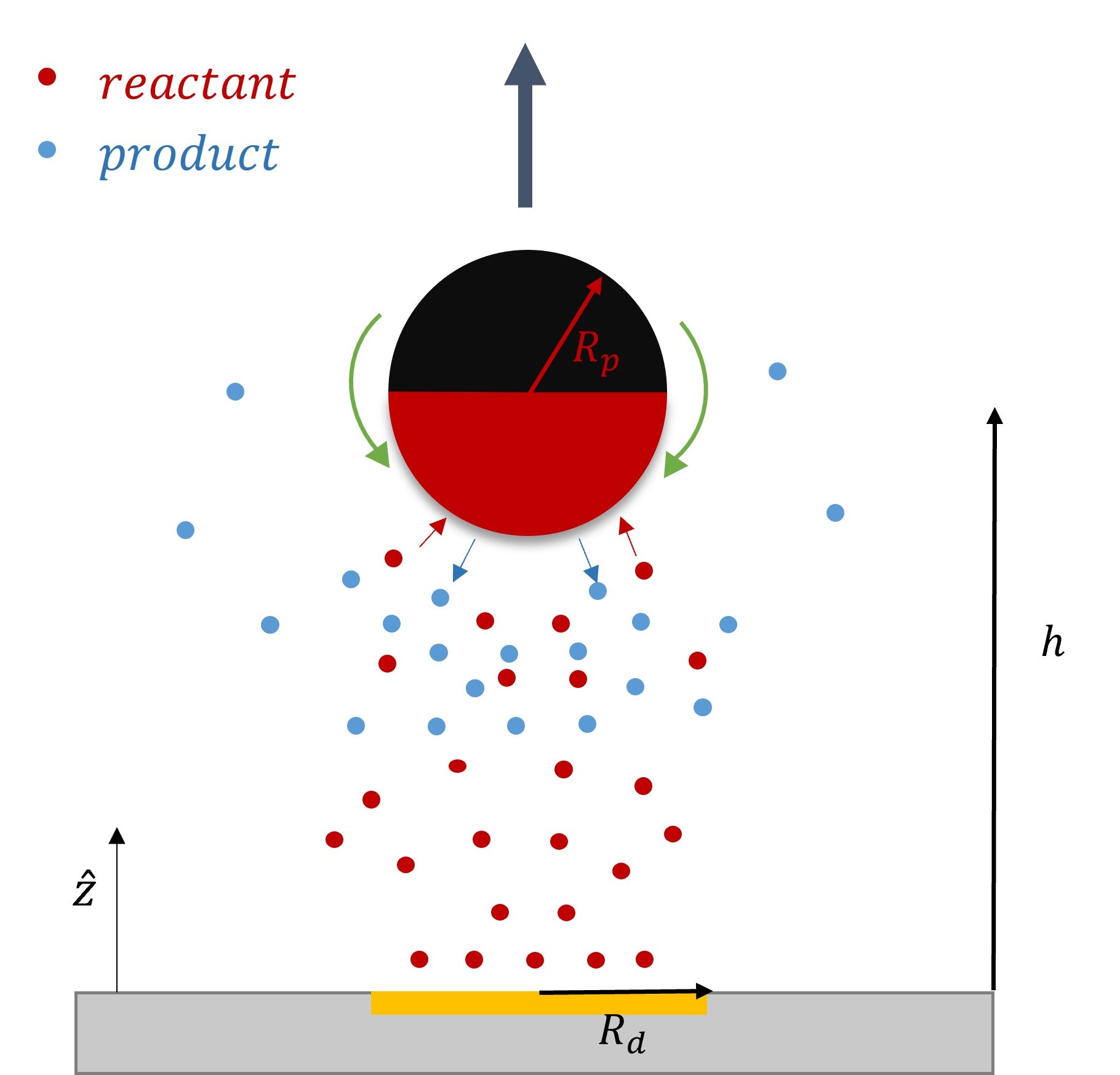}
\caption{Illustration of a spherical Janus particle (JP) and the patch-activated self-diffusiophoresis process. A Janus particle of radius $R_p$ is suspended in a liquid solution near a planar wall (grey). A circular patch with radius $R_d$ (orange) is located on the wall. The patch continuously releases reactant molecules (small red dots) into the solution. Reactant molecules are converted (red and blue arrows) into product molecules (small blue dots) at the catalytic cap of the Janus particle (red). The resulting gradient in product molecule concentration along the surface of the particle leads to an interfacial flow (green arrows), and therefore directed motion (black arrow). }
  \label{Figure 1}
\end{figure}

\section{Model}
Figure \ref{Figure 1} schematically depicts the key components of our model. A planar wall confines the suspension containing the Janus particle to the upper half space. At a region (the patch) positioned on the planar wall, reactant molecules are released into the surrounding solution. These rapidly diffuse throughout the solution; upon contact with the catalytically active side (the red hemisphere in \fref{Figure 1}) of the Janus particle (JP), they are converted, with certain probability (encoded by the rate of the chemical catalytic reaction), into product molecules. The chemical reaction can be simplified as $R \rightarrow a \cdot P$, with $R$ and $P$ representing a generic reactant and product, respectively, and $a$ as the stoichiometric factor; in the following we set $a=1$, for simplicity. Since the reaction exclusively occurs on one side of the JP, it leads to inhomogeneities in the spatial distribution of product and reactant molecules around the JP. We assume that the motion of the JP emerges through a phoretic response \textit{solely} to the gradients in the density of product molecules and further assume that this can be summarized by a phoretic actuation (active slip velocity, depicted schematically in \fref{Figure 1} by the green arrows) of the suspension at the surface of the particle, as in the classic framework by Anderson \cite{anderson1989colloid}. This gives rise to bulk hydrodynamic flow and motion of the particle. Under the typical assumptions of low Reynolds number flow of the solution and overdamped motion of the colloidal particle, which are very reasonable approximations in many experiments, in the absence of an external driving field acting on the particle, the force and torque induced by the interfacial flow are instantaneously balanced by translational and rotational hydrodynamic drag on the particle.

From a mathematical standpoint, the model described above  means determining a chemical field, which consists of the spatial distributions of reactant and product molecules, and a hydrodynamic field induced by the active actuation at the surface of the particle, as well as the overdamped motion of the particle. These points will be addressed in the next subsections. Before proceeding, here we discuss several approximations, already alluded to in the previous paragraph, which are suggested by the typical experimental studies of micrometer sized colloids in aqueous suspensions (these seem to be also the most suitable set-ups for realizations of the model system considered by us.) These will allow significant simplifications of the mathematical description.

First, for systems involving suspensions with density and viscosity similar to that of water, particles with sizes typically in the order of micrometers, and flow and particle velocities typically in the order of micrometers per second, viscous forces dominate over inertia. This is quantified by the Reynolds number $\textrm{Re} \equiv \frac{\rho |U| R_p}{\mu}$, where $\rho$ is the density of the fluid, $|U|$ a characteristic velocity (e.g., the translational velocity of the particle), and $\mu$ is the dynamic viscosity. Using  $R_p \sim 1 \;\mu\textrm{m}$,  $|U|  \sim 1 \;\mu\textrm{m} / \textrm{s}$, $\rho \sim 1000\;\textrm{kg}/\textrm{m}^3$, and $\mu \sim 10^{-3}\; \textrm{Pa} \cdot \textrm{s}$ renders $\textrm{Re} \sim 10^{-6} \ll 1$. Accordingly, the hydrodynamics is governed by the Stokes equation, and, consistently, the motion of the particle is in the overdamped regime.

Second, by the Stokes-Einstein relation, the diffusion of the molecularly sized reactant and product species in aqueous solutions is fast, implying that the transport of the reactant and product species by diffusion dominates over that by  advection. This assumption is substantiated by the value of the P\'{e}clet number (Pe), which compares the time scales of diffusion and advection over a lengthscale comparable to that of the particle size: the same numbers as above render $D \sim 10^{-9} \; \textrm{m}^2 / \textrm{s}$ for {\AA}-sized molecules and thus $\textrm{Pe} =  |U|  R_p / D \approx 10^{-3} \ll 1$. Consequently, the advection term can be neglected in the continuity equation for the product. This is a major simplification because it effectively decouples the equations governing the chemical field from the hydrodynamics of the solution. 

\subsection{Chemical Field}

The fast diffusion of the molecular species, $\textrm{Pe} \ll 1$, additionally justifies the assumption that, on the time scale of particle motion, the chemical field is quasi-relaxed to the steady-state distribution corresponding to the instantaneous position of the particle. Furthermore, we also assume that the molecules exhibit ``ideal-gas-like'' behavior, implying that the distribution of each species remains unaffected by the presence of other molecules. 

Given these considerations, the reactant chemical field  $c_{r}$ is modeled by the Laplace equation, 
\begin{subequations}
\label{eq:bvp_Cr}
    \begin{equation}
    \label{eq:Lap_Cr}
        \mathbf{\nabla} ^{2} \; c_{r} \; = \; 0\,,
    \end{equation}
subject to the boundary conditions (BC):\newline 
- at infinity
\begin{equation}
    \label{eq:BC_inf_Cr}
    \lim_{{|\mathbf{r}| \to \infty}} c_r = 0\,,
\end{equation}
meaning that no reactant molecules are present at a considerable distance from the patch; \newline
- release of reactant molecules at the patch at the wall (the plane $z = 0$): 
\begin{equation}
    \label{eq:BC_wall_Cr}
    - \; D_{r}  \; [\mathbf{\nabla} c_{r} \cdot  \mathbf{\hat{z}}]\vert_{z=0} = Q K(\mathbf{r})\vert_{z=0}\,,
    \end{equation}
 where $D_{r}$ denotes the reactant diffusion coefficient, $\mathbf{\hat{z}}$ the normal vector of the wall defined to point into the fluid,  $K(\br)$ a function that specifies the shape of the patch and characterizes the spatial distribution of activity, which equals 1 at the patch and 0 outside, and $Q$ (with units of m$^{-2}$ s$^{-1}$) denotes the number of reactant molecules released per unit area, per unit time. We assume $Q$ to be constant in time and space over the area occupied by the patch; \newline
- a sink at the active side, S$_{A}$ of the JP surface S where the conversion of reactant into product is promoted, and a reflective surface at the inert side S$_{I}$ of the JP surface, i.e., 
\begin{equation}
    \label{eq:BC_JP_Cr}
     -D_{r} \; [ \; \mathbf{\nabla} c_{r} \cdot \mathbf{\hat{n}} \; ] \; \vert_{S}= - F \; [f(\mathbf{r})\; c_r] \; \vert_{S}.
\end{equation}
We adopt first-order reaction kinetics, with the sink-strength directly proportional to the concentration of the reactant species, c$_r$. The activity function $f(\br)$ characterizes  the distribution of the reaction over the JP surface (\textit{f}(\textbf{r}) = 1 on S$_{A}$ and 0 on S$_{I}$), and the parameter \textit{F} (units of m s$^{-1}$) accounts for the rate of reactant conversion.  
\end{subequations}

As discussed, we consider in this study only the case of patches with axisymmetric shapes, and thus we write $K(\textbf{r}) = K(r)$ ($z = 0$, implicitly) and define the origin of the coordinate system at the center of the patch. The solution to the diffusion problem outlined above is more conveniently expressed as the superposition of a background field produced by the patch in the absence of the JP,  and a ``disturbance'' field,  $ c_{r}= \; c_{r}^{b} \; + \; c_{r}^{d} $. The background field obeys the Laplace equation, the boundary condition at infinity, \eq{eq:BC_inf_Cr}, and accounts for the boundary condition at the wall \eq{eq:BC_wall_Cr}.  This problem can be solved analytically using the separation of variables and the Fourier-Bessel transform, \cite{Sneddon_book,Gaspar_2019} yielding the solution:
\begin{equation} 
\label{e7}
c_{r}^{b} (\mathbf{r})  =  \int_{0}^{\infty} A(\xi) \; e^{-\xi z} \; J_{0}(\xi  r) \; d\xi,
\end{equation}
where $A(\xi)$ represents the Fourier-Bessel spectrum, and  $J_{0,1}$ denotes Bessel functions of the first kind. Since $\xi$ has dimensions of inverse length, note that $A(\xi)$ has dimensions of concentration times length. It can be found by applying the boundary conditions and noting that the Bessel functions $J_0$ obey the orthogonality relation:
\begin{equation}
    \int_0^\infty J_0(\xi r) \; J_0(\xi' r) \; r \; dr = \frac{\delta(\xi - 
    \xi')}{\xi}, \nonumber
\end{equation}
where $\delta(\xi - \xi')$ is the Dirac delta function. The boundary condition on the wall and the orthogonality relation give the Fourier-Bessel spectrum:
\begin{equation}
A(\xi') = \frac{C_0}{R_p} \int_0^{\infty} dr \; r \; J_0( \xi' r) \; K(r),
\end{equation}
where we define the characteristic concentration $C_0 \equiv Q R_{p} / D_r$.
The Fourier-Bessel spectrum can be introduced into equation \ref{e7}, completing the calculation of the background reagent chemical field as:
\begin{equation}
\label{e4}
    c_{r}^{b} (r,z) = \frac{C_0}{R_p}  \int_{0}^{\infty} d\xi \left[ \; \int_{0}^{\infty} dr' \; r' \; J_{0}(\xi r') \;  K(r')\right] e^{-\xi z} \; J_{0} (\xi r).
\end{equation}
The integral within the square brackets in \eq{e4} can be calculated exactly for a circular patch of radius $R_d$:
\begin{equation}
\int_{0}^{{R_d}} dr \; r \; J_{0}(\xi r) = \frac{R_d}{\xi} \; J_{1}(\xi R_d). \nonumber
\end{equation}
This leads to: 
\begin{equation}
\label{e5}
    c_{r}^{b} \; (r,z) = C_{0} \frac{R_d}{R_p} \;  \int_{0}^{\infty} \; d\xi \; \frac{1}{\xi} \; J_{1}(\xi R_d) \; e^{-\xi z} \; J_{0} \; (\xi r).
\end{equation}
Note that the factor of $\frac{R_d}{R_p}$ is due to our choice to use the particle size $R_p$ as a characteristic length scale in the definition of $C_0$. The combination ${C_{0} \frac{R_d}{R_p}} = \frac{Q R_d}{D_r}$ gives a characteristic concentration $C_c \equiv \frac{Q R_d}{D_r}$ sourced by a circular patch of radius $R_d$.

Anticipating an extension of the type of geometries analysed as a test for the theoretical far-field, point-particle predictions (see the following Section), we note here that from the solution for the background reactant field sourced by a circular patch, we can easily obtain the solution for the background field sourced by, e.g., a ring-shaped patch. Specifically, the ring-shaped patch has $K(r) = 1$ for $R_i < r < R_o$, and $K(r) = 0$ elsewhere. Due to the linearity of Laplace's equation, we can obtain the solution for a ring-shaped patch by subtracting the background field for a circular patch of radius $R_d = R_i$ from the background field for a circular patch of radius $R_d = R_o$.

In terms of the --- now known --- background density field $c_{r}^{b}$, the diffusion problem for the disturbance field is formulated as follows. The field $c_{r}^{d}$ obeys the Laplace equation, the boundary condition at infinity, \eq{eq:BC_inf_Cr}, and --- owing to $c_{r}^{b}$ already fulfilling \eq{eq:BC_wall_Cr} --- a zero normal current (reflective) boundary condition at the wall. The concentration gradient along the direction normal to the JP surface assumes the form: 
\begin{equation}
   \label{Equation 1}
   [  \nabla c_{r} \; \cdot \mathbf{\hat{n}} \; ] \; \rvert_{{S}} \; = \; [ \; \nabla  c_{r}^{b} \cdot \mathbf{\hat{n}} \; + \; \nabla c_{r}^{d} \; \cdot \mathbf{\hat{n}} \; ] \; \rvert_{{S}}\,,  \nonumber
\end{equation}
which, when combined with \eq{eq:BC_JP_Cr}, renders the BC at the JP surface
\begin{equation}
\label{chem2}
 \; \left[ \; \nabla \; c_{r}^{b} \cdot \mathbf{\hat{n}} \; + \; \nabla \; c_{r}^{d} \; \cdot \mathbf{\hat{n}} \; \right] \; \biggr\vert_{{S}} \; = \;  \frac{F \; }{D_{r} } \; \left[f(\mathbf{r}) \; \left( \; c_{r}^{b} \; + \; c_{r}^{d} \; \right)\right] \; \biggr\vert_{{S}} \,,
\end{equation}
or, in a non-dimensional form (with $\tilde \br = \br/R_p$ and $\tilde{c} = c/C_0$),
\begin{equation}
    \label{chem3}
    \left[ \; \tilde{\nabla} \tilde{c}_{r}^{b} \cdot \mathbf{\hat{n}} \; + \; \tilde{\nabla}\tilde{c}_{r}^{d} \cdot \mathbf{\hat{n}} \; \right] \; \biggr\vert_{{S}} =  \textrm{Da} \; \left[f(\tilde{\mathbf{r}}) \; \left( \; \tilde{c}_{r}^{b} \; + \; \tilde{c}_{r}^{d} \;\right)\right] \; \biggr\vert_{{S}}
\end{equation}
in terms of the dimensionless Damköhler number defined as 
\begin{equation}
\mathrm{Da} := F R_p / D_r \equiv F C_0/Q\,.   
\end{equation}
The Damköhler number contrasts the reaction rate (in the numerator) with the diffusion rate (in the denominator); alternatively, it can be seen, by the second expression, as comparing the relative rates of supply of fuel by the patch (the denominator) and consumption of fuel by the particle (the numerator).  At low Da values, the rate of molecular diffusion surpasses the rate of chemical reactions. In this regime, reactions become the limiting factor. Fuel is plentifully available for the JP, but the product concentration, and therefore the particle self-propulsion velocity, is limited by the slow catalytic reaction.  In contrast, at high Da values, chemical reactions prevail over molecular diffusion, swiftly converting reactant molecules into fuel and rendering diffusion the limiting factor. As a result, a fuel-depleted boundary layer forms near the JP. When $\textrm{Da} \sim 1$, the rates of both diffusion and reaction processes are comparable, ensuring a balanced interplay without a dominant mechanism. The implications of these varying scenarios on the JP's behavior, especially in terms of the reactant field and the particle velocity, merit further examination and are discussed in subsequent sections. 

Turning now to the distribution of product, we recall that at the JP surface the R-molecules are converted into P-molecules.  At steady state, the product chemical distribution is the solution of the Laplace equation,
\begin{subequations}
\begin{equation}
    \label{eq:Lap_Cp}
    \nabla^{2}c_{p} \;  =  \; 0\;,
\end{equation}
vanishing at infinity
\begin{equation}
    \label{eq:BC_inf_Cp}
    \lim_{{|\mathbf{r}| \to \infty}} c_p = 0\;,
\end{equation}
and obeying the BC at the JP particle surface
\begin{equation} \label{e10}
[\tilde{\mathbf{\nabla}} \tilde{c}_{p} \cdot \; \mathbf{\hat{n}} ] \; \vert_{{S}} 
\; = \;
- \textrm{Da} \; \frac{D_r}{D_p} \; 
\left[f(\mathbf{r}) \; \tilde{c}_{r}\right]   \; \vert_{{S}} 
\end{equation}
For simplicity, in the followings we assume equal diffusion coefficients, $D_{p}  = D_{r}$. 

\end{subequations}

\subsection{Fluid actuation by active phoretic slip}

As discussed, the interfacial flow driven by the spatially inhomogeneous distribution of the product concentration will be accounted in our model by an active phoretic slip velocity $\mathbf{v}_{s}$, which defines the relative velocity between the particle and the fluid. For this we assume the classic expression \cite{derjaguin,anderson1989colloid} 
\begin{equation}
    \label{eq:vs}
    \bv_s(\br) = - b(\br) \; \nabla_{||} \; c_p\;,
\end{equation}
where  $\nabla_{||} \; c_{p}$ defines the surface gradient of the product concentration, and $b$ defines the phoretic mobility, also known as the surface mobility, with units of m$^5$/s. 

The latter  is a material-dependent parameter that describes the effective intermolecular interaction between the product and the colloid surface. The value of this parameter quantifies the strength of intermolecular forces, and the sign indicates whether these interactions are attractive (positive sign) or repulsive (negative sign). Moreover, the phoretic mobility can be either \textbf{uniform} or\textbf{ non-uniform} over the particle surface. Typically, as the two sides of the Janus particle are composed of different materials, their distinct chemical properties result in separate and independent interactions with the product molecules. Consequently, the parameter ``$b$'' assumes distinct values $b_{c} \, b_0$ and $b_{i} \, b_0$,  on the cap and inert side, respectively.  Here, $b_c$ and $b_i$ are dimensionless quantities, while $b_0 > 0$ is a characteristic surface mobility. When considering the materials commonly utilized in the design of self-phoretic Janus particles, such as Pt and SiO$_{2}$, experimental evidence suggests that the ratio $b_{i}/b_{c}$  typically falls within the range of 0 to 0.3 \cite{katuri2021inferring}. This parameter holds particular significance in this problem as it can be adjusted to achieve specific motion dynamics. 

Finally, we note that from the form of the slip velocity, we can define a characteristic velocity $V_0 \equiv b_0 C_0 / R_p$. Substituting our expression for $C_0$, we obtain 
\begin{equation}
    \label{eq:def_V0}
    V_0 = b_0 Q / D_r\,.
\end{equation} 

\subsection{Hydrodynamic field} 
We assume that the flow is incompressible, giving the condition 
\begin{equation}
    \label{eq:incomp}
\nabla \cdot \mathbf{u} = 0\;,
\end{equation}
where  $\mathbf{u}(\br)$ is the velocity of the fluid flow (the hydrodynamic field).  As noted, for  $\textrm{Re} \ll 1$ the flow is the solution of the Stokes equation, 
\begin{subequations}
\begin{equation}
    \label{eq:Stokes}
    \nabla \cdot \bm{\sigma} = 0\;;
\end{equation}
for a Newtonian fluid, $\bm{\sigma} = -p \; \mathbf{I} + \mu \; [ \; \nabla \mathbf{u} + \nabla \mathbf{u}^T \;]$, where $p(\mathbf{r})$ is the pressure. This is subject to BCs of quiescent flow far from the colloid,
\begin{equation}
    \label{eq:BC_inf_flow}
    \mathbf{u} \; (|\mathbf{r}| \rightarrow \infty) = 0\;,
\end{equation}
and of a prescribed slip at the surface of the particle,
\begin{equation}
    \label{eq:BC_JP_flow}
    \mathbf{u(r)}\vert_S = \left[\mathbf{V} \; + \;  \bm{\Omega} \; \times \; (\mathbf{r} - \mathbf{r}_p) +  \mathbf{v}_s(\mathbf {r})\right]_S\;,
\end{equation}
where $\mathbf{V}$ and $\bm{\Omega}$ represent the translational and angular velocities of the Janus particle, respectively, 
$\mathbf{v}_s(\mathbf{r})$ is the slip velocity, and $\mathbf{r}_p$ is the position of the particle center. 
\end{subequations}

To close the system of equations (recall that $\bV$ and $\bOmega$ are unknown at this point), we consider the net force and the net torque acting on the colloid, defined as \textbf{F}$_{net}$  and the \textbf{T}$_{net}$ respectively. Since $\textrm{Re} \ll 1$, these instantaneously vanish: $\textbf{F}_{net} \approx 0$  and $\textbf{T}_{net} \approx 0$.  Both can be seen as the sum of hydrodynamic and external contributions so that:  \textbf{F}$_{net}$ = \textbf{F}$_{h}$ + \textbf{F}$_{ext}$ and  \textbf{T}$_{net}$ = \textbf{T}$_{h}$ + \textbf{T}$_{ext}$. The hydrodynamic terms, \textbf{F}$_{h}$ and \textbf{T}$_{h}$, are exerted by the fluid on the particle. On the other hand,  \textbf{F}$_{ext}$ and \textbf{T}$_{ext}$ refer to external forces and torques acting on the system. In this study, we assume that there are no external forces or torques on the system.  Consequently, the net force becomes: 
\begin{subequations}
\begin{equation}
\label{F1}
  \mathbf{F}_{net} 
  =   \int\limits_S \; \bm{\sigma} \cdot \; \mathbf{\hat{n}} \; dS  = \; 0.
\end{equation}
Similarly, the net torque becomes: 
\begin{equation}
\label{F2}
  \mathbf{T}_{net} 
  =  \int\limits_S \; (\mathbf{r} - \mathbf{r}_p)  \; \times \; \bm{\sigma} \; \cdot \; \mathbf{\hat{n}}\; dS  = \; 0\,.
\end{equation}
\end{subequations}

At this stage, the hydrodynamic problem has been fully characterized and the analysis can be simplified by introducing the Lorentz Reciprocal Theorem, which connects the slip velocity and the translational and rotational velocities, leading to:\cite{poehnl2021phoretic}
\begin{equation}
    \label{F4}
    \begin{split}
    \mathbf{V} \; \cdot \;  \mathbf{F}_{a}  \; +  \; \bm{\Omega}  \; \cdot \; \mathbf{T}_{a} = - \;
    \int_{S} \mathbf{v_{s}}  \cdot  \; \bm{\sigma}_{a} \; \cdot \; \mathbf{\hat{n}} \; dS.
    \end{split}
\end{equation}
Here, the subscript ``a'' denotes an auxiliary hydrodynamic problem that has the same geometry as the hydrodynamic problem formulated above, but with different boundary conditions for the fluid velocity. The auxiliary force, torque, and stress denoted as $\mathbf{F}_{a}$, $\mathbf{T}_{a}$, and $\bm{\sigma}_{a}$, respectively, depend on the specific choice of the auxiliary problem. Judicious selection of auxiliary problem(s) with known or calculable solutions for $\mathbf{F}_{a}$, $\mathbf{T}_{a}$, and $\bm{\sigma}_{a}$ allows specification of a linear system for the unknown components of $\mathbf{V}$ and $\bm{\Omega}$. (Some may be known to be zero \textit{a priori} by symmetry.) Further details are available in Ref. \citenum{poehnl2021phoretic}.

\section{Results} 
In the following, we assume an axisymmetric configuration of the patch and the particle, as schematically illustrated for a sphere by Fig. \ref{Figure 1}. We aim to understand whether the particle will strictly move towards or away from the patch, or whether, under some conditions, the particle can have a motionless ``hovering'' state at a fixed height above the patch. Additionally, for hovering states, we seek to understand the dependence of the hovering height on the system parameters ($\textrm{Da}$, $b_i$, and $b_c$), as well as the stability of the hovering state against perturbations in the particle height.

In the preceding, we did not assume any particular shape of the Janus particle. In the following, we will mainly consider spherical particles of radius $R_p$ that are half-covered by catalyst. We will also consider prolate spheroidal particles with semi-major axis length $R_p$. The semi-minor axis length is quantified by the aspect ratio $r_e$ as $r_e \, R_p$, where $0 < r_e \leq 1$. Both spherical and prolate spheroidal particles are assumed to have  axisymmetric catalyst coverage, and we will also consider the dependence of hovering on the particle shape, i.e., the value of $r_e$.

\subsection{Concentration Fields }

As described above, the determination of the reactant and product concentration fields is independent of, and logically precedes, the determination of the particle velocity. Therefore, we first consider these fields for a particle near a wall.  The background reactant field $c_r^b$ is obtained by numerical integration of Eq. (\ref{e5}). We solve for the concentration fields $c_r^d$ and $c_p$ using two approaches. Semi-analytically, we solve for the fields using bipolar coordinates, as detailed in Appendix \ref{sec:App_bipolar}. Numerically, we use the boundary element method (BEM) \cite{pozrikidis2002practical}.

\begin{figure}
    \centering
    \includegraphics[width=\linewidth]{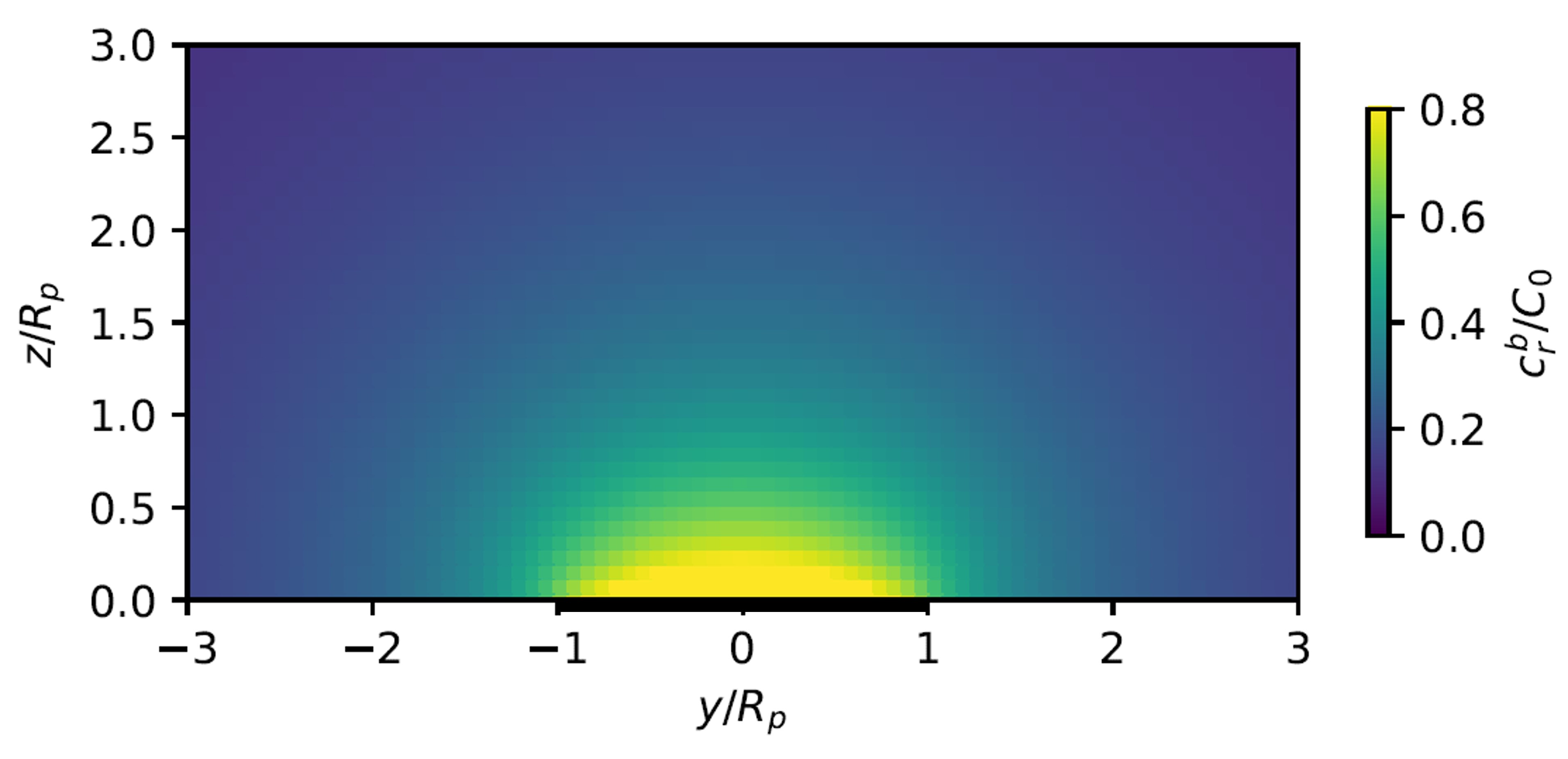}
    \caption{Background reactant concentration $\tilde{c}_r^b$ sourced by a circular patch with radius $R_d = R_p$. The patch is shown as a thick black line.}
    \label{fig:cbr}
\end{figure}

We briefly discuss the reactant background field, shown in Fig. \ref{fig:cbr} for a circular patch with $R_d = R_p$. In the bulk liquid near the patch ($0.1 < z/R_p < 1$ and $|y|/R_p < 1.5$), there is strong spatial variation of the concentration. In  very close vicinity of the patch, $z \approx 0$ and $|y|/R_p < 1$, the reactant field is approximately uniform in the lateral direction. At this close distance, the patch resembles a planar wall that uniformly releases reactant. Since the only length scale in determination of the background field for a circular patch is $R_d$, the values of the dimensionless concentration for other values of $R_d$ can be easily inferred by rescaling of the coordinate axes and the dimensionless concentration by $R_d/R_p$ (see the prefactor in Eq. (\ref{e5}).)

\begin{figure*}
\centering
 \includegraphics[height=10cm]{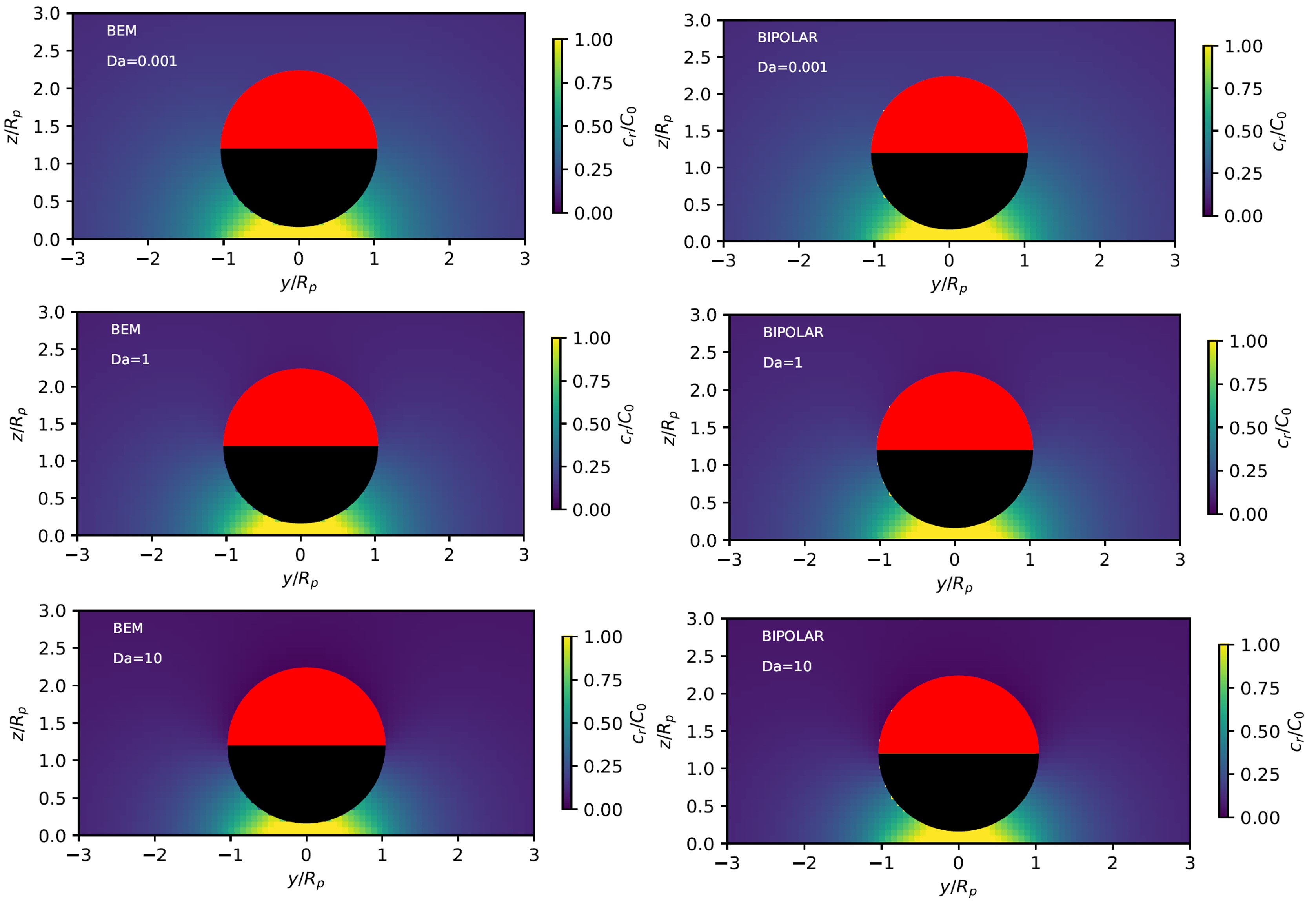}
 \caption{Concentration of the chemical reactant for a cap-up sphere near a chemical patch for various values of the Damköhler number $\textrm{Da}$. The particle is located at $h/R_p = 1.2$, and the patch size is $R_d = R_p$.}
 \label{fig:conc_reactant}
\end{figure*}
\begin{figure*}
\centering
 \includegraphics[height=10cm]{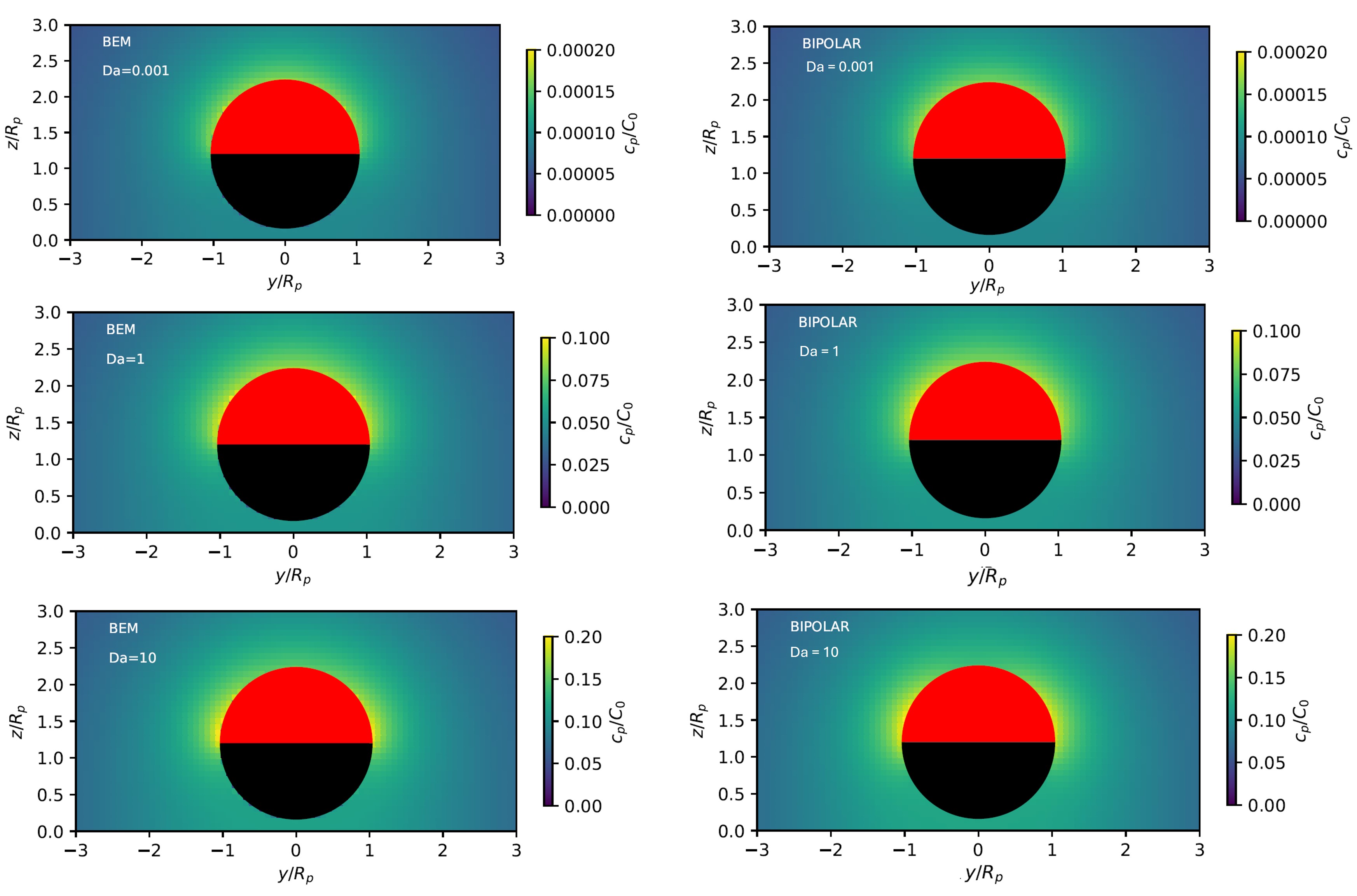}
 \caption{Concentration of the chemical product for a cap-up sphere near a chemical patch for various values of the Damköhler number $\textrm{Da}$. The particle is located at $h/R_p = 1.2$, and the patch size is $R_d = R_p$.}
 \label{fig:conc_product}
\end{figure*}

Fig. \ref{fig:conc_reactant} shows the reactant concentration for $h/R_p = 1.2$ and $\textrm{Da} = 0.001, 1, 10$. For the low value of $\textrm{Da}$, diffusion is the predominant process. A particular concentration, $c_r/C_0 = 1$ (yellow tone) reaches a distance on the wall ($z = 0)$ equal to  $y/R_p = 0.8$. In the vicinity of the catalytic ``pole'' of the particle, located at $z/R_p = 2.2$, the reactant concentration $c_r/C_0 \approx 0.15$ (bluish color). At $\textrm{Da} = 1$, the two processes (reaction and diffusion) assume equal importance.  At high $\textrm{Da}$, the reactant is quickly consumed by the catalytic cap, and therefore is less able to diffuse into the surrounding solution. Here, $c_r/C_0 = 1$ reaches only $y/R_p = 0.5$ on the wall. Moreover, the reactant concentration is nearly exhausted in the vicinity of the catalytic pole ($c_r/C_0 \approx 0.01$, purple color.) In all cases, the results obtained numerically, with the BEM, and semi-analytically, using bipolar coordinates, show excellent quantitative agreement. 

Fig. \ref{fig:conc_product} shows the product concentration for the same height and values of $\textrm{Da}$ as in Fig. \ref{fig:conc_reactant}. For the smallest value of $\textrm{Da}$, $\textrm{Da} = 0.001$, the concentration is highest near the catalytic pole. The maximum of the product concentration being located at the catalytic pole is a typical scenario for catalytic Janus spheres \cite{michelin2014phoretic}. Over the cap, the spatial gradient of concentration is directed from the ``equator'' to the pole, with a positive component in the z-direction. Over the inert region of the particle, the spatial gradient is from the pole to the equator; again, the z-component is positive. Recalling that $\mathbf{v}_s(\mathbf{r}) = -b(\mathbf{r}) \nabla_{||} c_p$, we expect that the slip velocity will have a positive z-component where $b$ is negative. Therefore, for $b_i < 0$, the inert region would drive swimming towards the wall. (Recall from Fig. \ref{Figure 1} that slip velocity induces translational motion in the opposite direction.) Likewise, for $b_c < 0$, the catalytic region would also drive swimming towards the wall. 

For the largest value of $\textrm{Da}$, however, we see an ``inversion'' effect. The concentration is highest near the ``equator'' of the particle. Thus, on the catalytic side, the spatial gradient of the concentration is directed from the catalytic ``pole'' to the ``equator,'' with a negative component in the z-direction.  Therefore, if $b_c < 0$, the slip velocity on the cap will be directed from the ``pole'' to the ``equator,'' and the catalytic cap would drive swimming \textit{away from} the wall. On the inert side, we still have a spatial gradient from the ``pole'' to the ``equator,'' and a slip velocity in the same direction for $b_{i} < 0$. Therefore, the inert side would still drive swimming towards the wall when $b_i < 0$, as in the low $\textrm{Da}$ case. 

\begin{figure}
    \centering
    \includegraphics[width=0.85\linewidth]{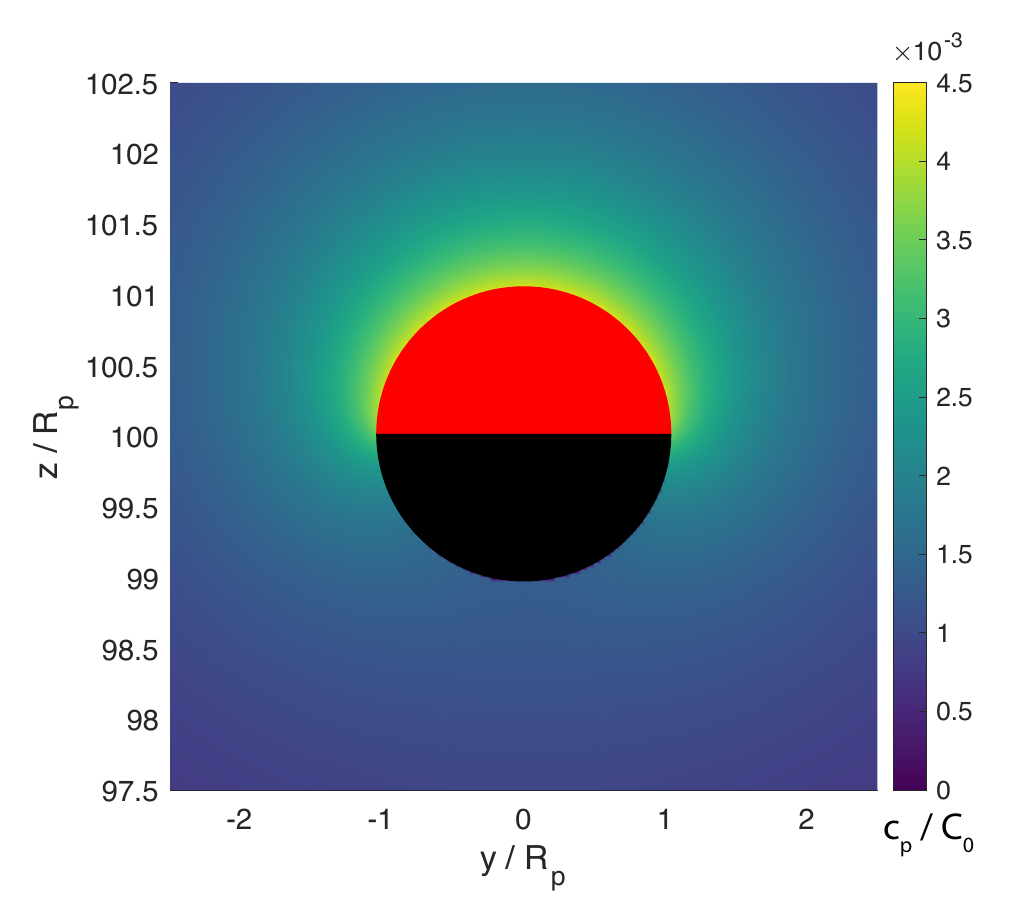}
    \caption{Concentration of the chemical product for a cap-up sphere located far away from the wall ($h/R_p = 100$) with $\textrm{Da} = 10$. The maximum of $c_p$ is at the catalytic ``pole,'' in contrast with the ``inversion'' effect seen for $\textrm{Da} = 10$ in Fig. \ref{fig:conc_product}.  The patch size is $R_d = R_p$, and the concentration was computed with the BEM.}
    \label{fig:Da_10_h_100}
\end{figure}

From the foregoing, it is clear that, for a given configuration of the particle and patch, the Damköhler number $\textrm{Da}$ can have a profound effect on the product concentration field. Increasing $\textrm{Da}$ for a particle at a given position can lead to a ``concentration-inversion'' effect. We can also show that, at a given value of $\textrm{Da}$, the inversion effect depends on proximity to the patch. In Fig. \ref{fig:Da_10_h_100}, we show the product concentration field for a sphere with $\textrm{Da} = 10$ that is far away from the patch ($h/R_p = 100$). The maximum of the concentration is at the catalytic pole, i.e., there is no inversion effect. Therefore, for $b_c < 0$, the catalytic cap would drive swimming towards the wall. 

Now we are in a position to qualitatively infer the existence of a ``hovering'' state. To simplify the following argument, suppose $b_i = 0$, such that the inert side of the particle does not contribute to the particle velocity. For $\textrm{Da} = 10$ and $b_c < 0$, the catalytic cap drives swimming away from the wall when the particle is at a height $h/R_p = 1.2$. For the same value of  $\textrm{Da}$ and the same $b_c < 0$, the catalytic cap drives swimming towards the wall when $h/R_p = 100$.  Thus, we expect that at some intermediate height $1.2 < h/R_p < 100$, the velocity of the particle will be zero. The particle will exhibit motionless ``hovering'' at this height (although it will continuously pump the surrounding fluid.) The following section will confirm this reasoning quantitatively. 

\subsection{Particle Velocity}

Having characterized how the reactant and product fields depend on the particle configuration and Damköhler number, we now turn to investigating the vertical motion of the particle, i.e., motion towards or away from the patch. In Appendix B, as an additional cross-check and validation for the boundary element method, --- which is used exclusively in the rest of the study because: (i) it straightforwardly handles other particle shapes as well as non-axisymmetric configurations, and (ii) computationally it is significantly faster, --- we provide a brief comparison of velocities obtained with the BEM and using bipolar coordinates. Throughout the rest of Section III, we use the BEM to determine both the concentration fields $c_r^d$ and $c_p$, as well as to solve the necessary auxiliary hydrodynamic problems as outlined in Section II.C. 

\subsubsection{Uniform Phoretic Mobility}

\begin{figure}[h]
\includegraphics[width=10cm]{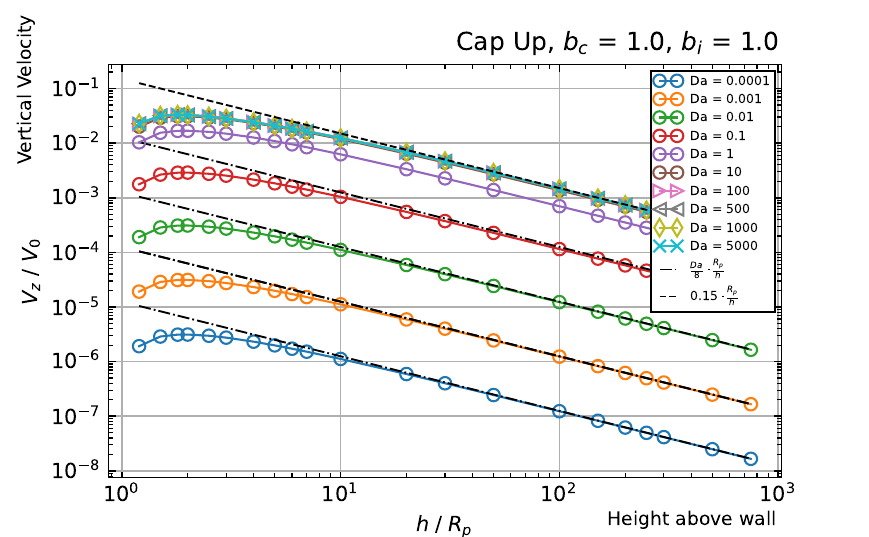}
\includegraphics[width=10cm]{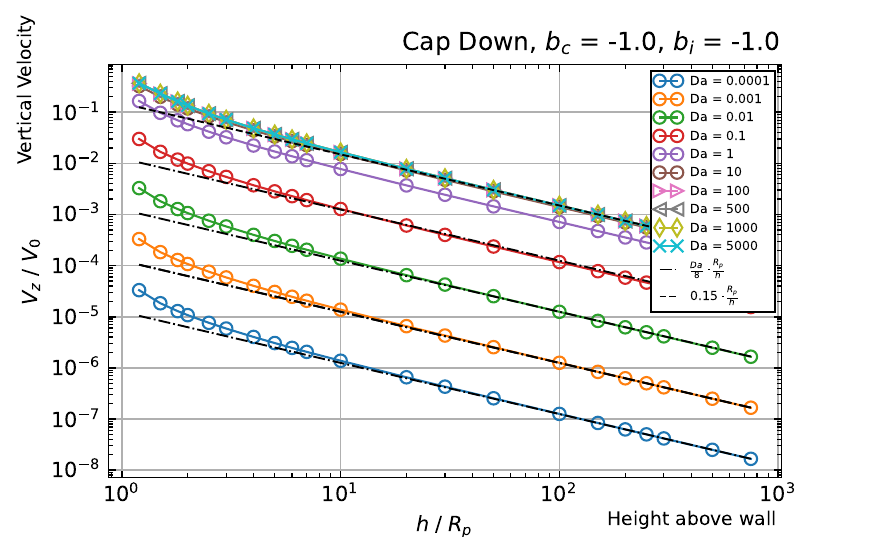}
    \caption{Dependence of the vertical velocity $V_z$ on the height $h$ for a sphere with uniform surface mobility near a circular patch. Predicted low \textrm{Da} and high \textrm{Da} scalings are given by dotted-dashed and dashed lines, respectively. The sign of the surface mobility is chosen to give a positive velocity. The patch has the same radius as the particle ($R_d/R_p = 1$).}
    \label{fig:uniform-mobility-sphere}
\end{figure}

We first consider the case of uniform phoretic mobility, $b_i = b_c$. As a starting point, we consider a spherical particle near a patch that has the same radius as the sphere ($R_d = R_p$). Figure \ref{fig:uniform-mobility-sphere} shows how the vertical velocity $V_z$ depends on height $h$ for different values of the Damk\"ohler number $\textrm{Da}$.  In the top panel, the sphere has a cap-up configuration, and in the bottom panel, the sphere is cap-down. In each panel, the sign of $b_i = b_c$ is chosen to give a positive value of $V_z$.  The log-log scaling of the plots reveals the dependence of $V_z$ on $h$ is described by a power law for $h/R_p \gg 1$. It also also evident that, for $\textrm{Da} \gg 1$ and $h/R_p \gg 1$, the $V_z$ vs. $h$ curves, which are calculated for different values of $\textrm{Da}$,  collapse onto a universal power law scaling.

The power law behavior can be recovered in the framework of a far-field, ``point-particle'' theory. We model the reactant background concentration as being due to a point source located at the origin:
\begin{equation}
\label{eq:pointsource}
c_r^b(\mathbf{r}) \approx \frac{2 A_{d} Q}{4 \pi D_r |\mathbf{r}|},
\end{equation}
where $A_d$ is the area of the patch. The factor of two in the numerator is due to confinement of the reactant in the upper half space by the wall. The background reactant concentration in the vicinity of the Janus particle can be expanded as follows:
\begin{equation}
\label{eq:crb_expansion}
c_r^b(\mathbf{r}) = c_r^b(\mathbf{r}_p) + \nabla c_{r}^b|_{\mathbf{r} = \mathbf{r_p}}  \cdot (\mathbf{r} - \mathbf{r}_p) + ..., 
\end{equation}
where the position of the particle is $\mathbf{r}_p = h \, \mathbf{\hat{z}}$. In this expansion, each successive term introduces another factor of $h/R_p$. Therefore, we generally expect that the first term, $c_r^b(\mathbf{r}_p)$, will provide the leading-order contribution to the velocity of the Janus particle. Focusing on the contribution of this term to $V_z$, we can interpret it as being due to an effectively uniform reactant field in the vicinity of the particle.

From Fig. \ref{fig:uniform-mobility-sphere}, it is apparent that for low values of $\textrm{Da}$, each $V_z$ vs. $h$ curve is associated with its own power law. Therefore, in the point-particle framework, we first consider the limit $\textrm{Da} \rightarrow 0$. In this limit, the consumption of the fuel by the particle is negligible, and we approximate $c_r(\mathbf{r}) \approx c_r^b(\mathbf{r})$. Moreover, in this limit, we expect to recover results from zeroth order kinetics, i.e., results for a Janus particle with a boundary condition $-D_p [\nabla c_p \cdot \mathbf{\hat{n}}] = \kappa$ on the catalytic cap, where $\kappa = F c_r^b(\mathbf{r}_p)$. It is known that a Janus particle in free space (unconfined solution), with its symmetry axis aligned with the z-axis,  will have velocity $V_z = \frac{b_0 \kappa }{D_p} \, \tilde{V}_{fs}$, where $\tilde{V}_{fs}$ is a dimensionless constant.  Therefore, we obtain the scaling 
\begin{equation}
V_z = \frac{A_d Q  F b_0 \tilde{V}_{fs}}{2 \pi D_r D_p h}.     
\end{equation}
Using the assumption that $D_p = D_r$, the definitions of $V_0$ and $\textrm{Da}$, and defining the dimensionless patch area $\tilde{A}_d = A_d / \pi R_p^2$, we obtain 
\begin{equation}
\label{eq:far_field_vz}
V_z / V_0 = \frac{\tilde{A}_d \, \textrm{Da} \: \tilde{V}_{\textrm{fs}}}{2} \, \frac{R_p}{h}.
\end{equation}
For a spherical Janus particle with uniform surface mobility, $\tilde{V}_{\textrm{fs}} = \pm 1/4$. The sign of $\tilde{V}_{\textrm{fs}}$ is determined by the orientation of the cap (up or down) and the sign of $b_i = b_c$. Values of $\tilde{V}_{\textrm{fs}}$ for spheroidal particles can be calculated analytically or numerically \cite{popescu2010phoretic,poehnl2021phoretic}.

\begin{figure}
    \centering
\includegraphics[width=10cm]{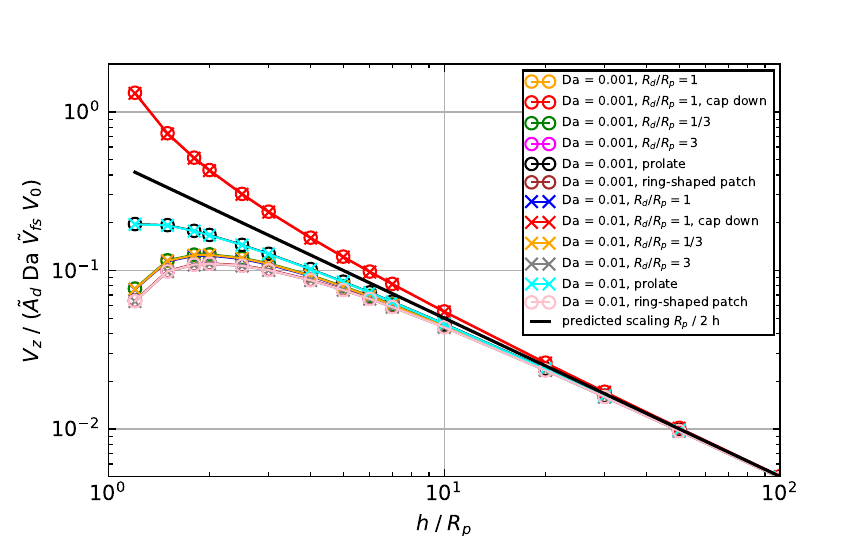}
    \caption{In the limit $\textrm{Da} \rightarrow 0$, the velocity $V_z$ as a function of height $h$ can be collapsed onto a universal curve, given by Eq. (\ref{eq:far_field_vz}), for various particle shapes, patch sizes, and patch shapes. Except where indicated in the legend, the shape of the particle is assumed to be spherical, the configuration is assumed to be cap-up, and the shape of the patch is assumed to be circular. The surface mobility is assumed to be uniform, with the sign chosen to give a positive velocity. For the prolate particle, $r_e = 1/3$, and $\tilde{V}_{fs}$ was determined numerically as $\tilde{V}_{fs} = 0.1247$. For the ring, the outer radius is $R_o/R_p = 2.5$ and the inner radius is $R_i/R_p = 1.5$. }
    \label{fig:low_Da_collapse}
\end{figure}

The predicted scaling, plotted as dashed-dotted lines in Fig. \ref{fig:uniform-mobility-sphere}, shows good agreement with the $\textrm{Da} \ll 1$ data. In order to emphasize the universality of this scaling for $\textrm{Da} \ll 1$, in Fig. \ref{fig:low_Da_collapse} we show data for $\textrm{Da} = 0.001$ and $\textrm{Da} = 0.01$ for cap-up and cap-down spheres activated by circular patches of different sizes $R_d$, a cap-up prolate particle activated by a circular patch, and a cap-up sphere activated by a ring-shaped patch. We exploit our scaling to collapse the data onto a universal curve.

Now we turn to the opposite limit, $\textrm{Da} \rightarrow \infty$. From our data in Fig. \ref{fig:uniform-mobility-sphere}, it is apparent that the function $V_z$ vs. $h/R_p$ for different values of $\textrm{Da}$ collapses onto a universal curve as $\textrm{Da} \rightarrow \infty$. In other words, the scaling for $V_z$ vs. $h/R_p$ loses dependence on $\textrm{Da}$ as $\textrm{Da} \rightarrow \infty$. We give arguments in Appendix A rationalizing this loss of dependence on $\textrm{Da}$. Moreover, in Appendix A, we argue that the prefactor of the scaling $V_z/V_0 \sim \tilde{A}_d R_p/h$ should be the dimensionless number $0.15$ for a sphere. This predicted scaling is shown on Fig. \ref{fig:uniform-mobility-sphere}, and has good agreement with the numerical data. 

\subsubsection{Non-Uniform Phoretic Mobility} 

\begin{figure}[!htb]
    \centering
    \includegraphics[width=10cm]{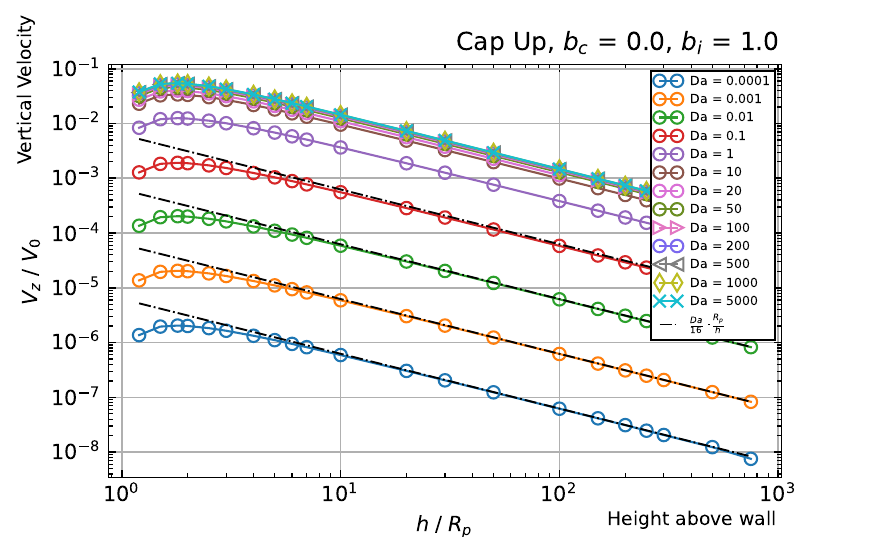}
    \includegraphics[width=10cm]{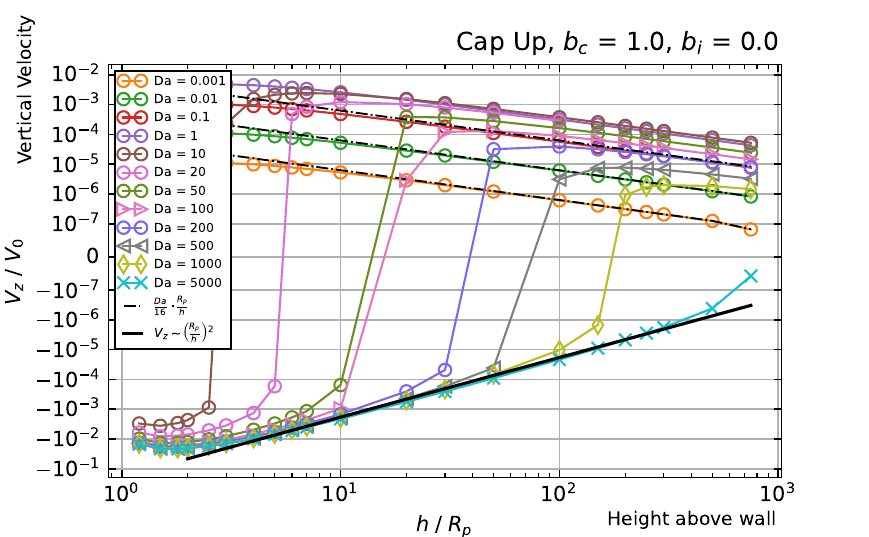}
    \includegraphics[width=10cm]{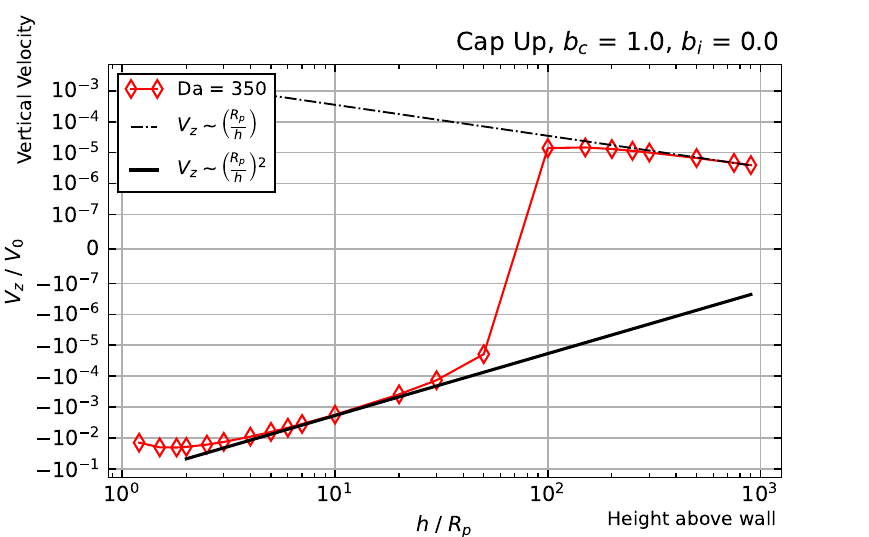}    
    \caption{(top and middle panels) Dependence of the vertical velocity $V_z$ on the height $h$ for a cap-up sphere with non-uniform surface mobility near a circular patch with $R_d = R_p$.  (bottom panel) Dependence of $V_z$ on $h$ for a cap-up sphere with $b_c = 1$, $b_i = 0$, and $\textrm{Da} = 350$. Crossover between two power law scalings is clearly visible. 
    }
    \label{fig:non-uniform-mobility}
\end{figure}
Taking advantage of the linearity of the hydrodynamic problem, for non-uniform phoretic mobility ($b_i \neq b_c$)  we can obtain the particle velocity by superposing the velocity obtained for $(b_i = 1, b_c = 0)$ and the velocity obtained for $(b_i = 0, b_c = 1)$. Specifically, the particle velocity is $V_z(h/R_p; \: b_i, b_c) = b_i \; V_z(h/R_p; \: b_i = 1, b_c = 0 ) + b_c \; V_z(h/R_p; \: b_i = 0, b_c = 1)$. Therefore, in what follows, we will focus on the two cases of $(b_i = 1, b_c = 0)$ and $(b_i = 0, b_c = 1)$, and then consider general values of $b_i$ and $b_c$.

In Fig. \ref{fig:non-uniform-mobility}, we show the results for a cap-up sphere for the cases of $b_c = 0$, $b_i = 1$ and $b_c = 1$, $b_i = 0$. In the first case (phoretic slip only on the inert side), we obtain a similar result as in the case of uniform mobility. Again, the low $\textrm{Da}$ curves are described by the scaling law given in Eq. (\ref{eq:far_field_vz}). In this case,  the value  of   $\tilde{V}_{fs}$ appearing in the low $\textrm{Da}$  scaling is $\tilde{V}_{fs} = 1/8$. We also obtain collapse for the $\textrm{Da} \gg 1$ curves, although for higher values of $\textrm{Da}$ than in the case of uniform mobility.  

Turning to the second case (phoretic slip only on the catalytic side), we obtain some intriguing behavior. Specifically, for $\textrm{Da} \gg 1$, we find that the curves have negative $V_z$ at low values of $h/R_p$, but switch over to positive $V_z$ at high values of $h/R_p$. Accordingly, each of these curves intersect $V_z = 0$ at some crossover height $h_c(\textrm{Da})$. These crossover points can be identified as hovering states. Since the slope of $V_z$ vs $h$ at the crossover height in Fig. \ref{fig:non-uniform-mobility} is positive, the states shown in that figure are unstable against perturbations of the particle height. However, changing $b_c = 1$ to $b_c = -1$ would invert the sign of the velocity $V_z$, and therefore change the sign of the slope, which  would make the hovering states stable against vertical perturbations.

Another intriguing aspect of $b_i = 0$, $b_c = 1$ can be observed for the $\textrm{Da} \gg 1$ curves. Specifically, they follow a $V_z \sim (R_p/h)^2$ power law at intermediate heights, and cross over to a $V_z \sim (R_p/h)$ power law at large heights.  The crossover behavior is especially apparent for $\textrm{Da} = 350$, shown in the bottom panel of Fig. \ref{fig:non-uniform-mobility}. The observed crossover between power laws can rationalized by the perturbation theory developed for $\textrm{Da} \gg 1$ in Appendix A. As shown there, in the limit $\textrm{Da} \rightarrow \infty$, and for a uniform background concentration of reactant, the product concentration on the cap would be uniform: $\tilde{c}_p^{(\infty)} = \tilde{c}_{r}^{b}(\mathbf{r}_p)$ on the cap, where the superscript $(\infty)$ indicates the $\textrm{Da} \rightarrow \infty$ limit. Therefore, for $b_c = 1$ and $b_i = 0$, there would not be a surface gradient of product concentration on the region of the particle surface with a non-zero phoretic mobility. The particle velocity would therefore be zero. However, in this limit ($\textrm{Da} \rightarrow \infty$), the particle can still self-propel if there is a background reactant \textit{gradient}. Since the background reactant field is sourced by a patch at the origin, the gradient in the vicinity of the particle, $\nabla c_{r}^b|_{\mathbf{r} = \mathbf{r}_p}$, decays as $\nabla c_{r}^b|_{\mathbf{r} = \mathbf{r}_p} \sim (R_p/h)^2$. Therefore, the contribution to velocity that is leading order in inverse powers of $\textrm{Da}$ is $V_z^{(\infty)} \sim (R_p/h)^2$.

\begin{figure}[!b]
    \centering
    \includegraphics[width=9cm]{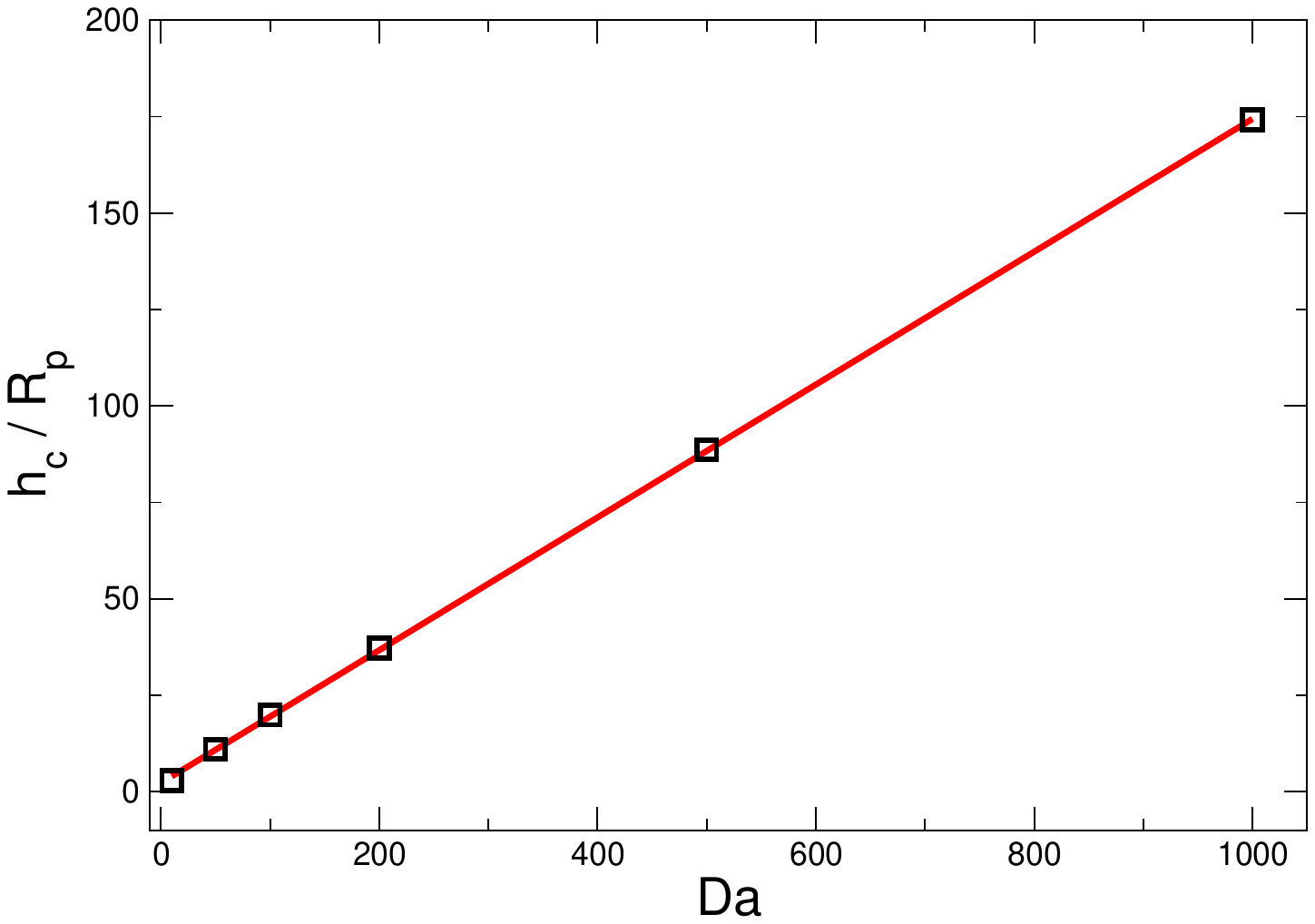}
    \caption{Crossover height $h_c$ as a function of \textrm{Da} for a spherical, cap-up Janus particle with $b_c = 1$, $b_i = 0$. The particle is near a circular patch with radius $R_d = R_p$. We approximate the crossover height by the value where the particle velocity $V_z$ is zero, as seen in the bottom panel of Fig. \ref{fig:non-uniform-mobility}. The red line in the figure is a linear fit $h_c/R_p = 0.172\; \textrm{Da} + 2.253$. As discussed in the text, we predict a linear dependence of $h_c/R_p$ on $\textrm{Da}$ using perturbation theory.}
    \label{fig:crossover}
\end{figure}
To complete the picture, we must explain the $V_z \sim R_p/h$ scaling at $h/R_p \gg 1$ and the crossover between the two power laws. Here, we note that there is a subleading $\mathcal{O}(\textrm{Da}^{-1}$) contribution to particle velocity for $\textrm{Da} \gg 1$. As detailed in Appendix A, the subleading contribution \textit{does} give particle motion in a uniform background concentration of reactant. Therefore, the subleading contribution to velocity goes as $V_z^{(1)} \sim  \frac{1}{\textrm{Da}} \left(\frac{R_p}{h}\right)$. Here, the superscript $(1)$ indicates that this contribution to the velocity $V_z$ is $\mathcal{O}(\textrm{Da}^{-1})$. For a fixed, finite value of $\textrm{Da}$, this $V_z^{(1)} \sim \frac{1}{\textrm{Da}} \left(\frac{R_p}{h}\right)$ scaling will eventually dominate the leading order $V_z^{(\infty)} \sim (R_p/h)^2$ scaling for sufficiently large $h$. The crossover height $h_c$ can be estimated from $\left(\frac{R_p}{h_c}\right)^2 \sim \frac{1}{\textrm{Da}} \left(\frac{R_p}{h_c}\right)$, or $h_c/R_p \sim \textrm{Da}$. In order to test this prediction, we show the crossover height as a function of $\textrm{Da}$ in Fig. \ref{fig:crossover}. We find that a linear relationship does indeed capture the dependence of $h_c/R_p$ on $\textrm{Da}$.

Our analysis explained the crossover phenomenon mathematically. A physical explanation can inferred from the ``inversion'' of the product concentration observed for $\textrm{Da} = 10$ in Fig. \ref{fig:conc_product}. When a particle is close to the patch, it experiences a strong reactant concentration gradient. Therefore, for the diffusion-limited regime ($\textrm{Da} \gg 1$), the region of the catalyst near the ``equator'' has a higher rate of reaction than the region near the ``pole,'' due to the much greater availability of fuel closer to the patch. This effect explains the importance of the reactant gradient $\nabla c_{r}^b$ in determining the particle velocity for $\textrm{Da} \gg 1$ and $h < h_c$. However, when the the particle is far away from the patch, as in Fig. \ref{fig:Da_10_h_100}, the reactant gradient is much weaker. Therefore, the product concentration field is qualitatively the same as for a particle in a uniform background reactant field, i.e., the maximum is at the pole.

\begin{figure} [!b]
\centering 
\includegraphics[width=10cm]{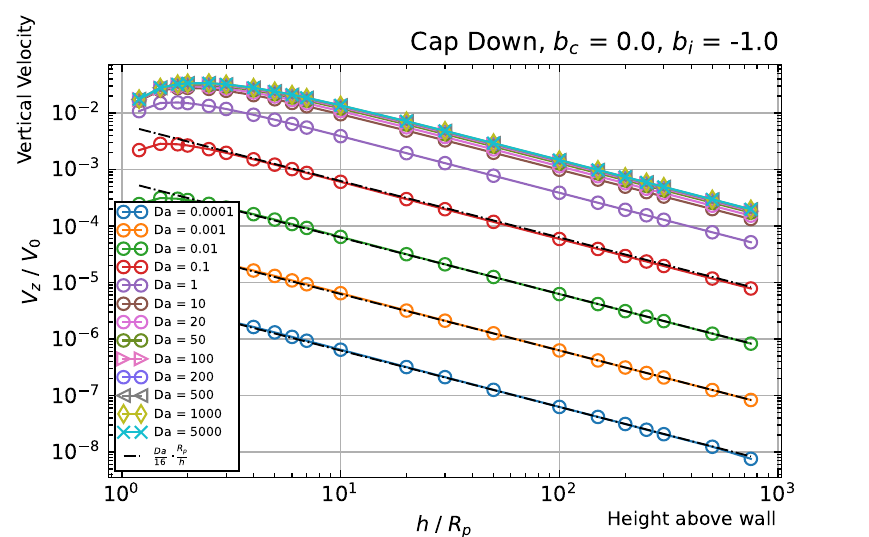}
\includegraphics[width=10cm]{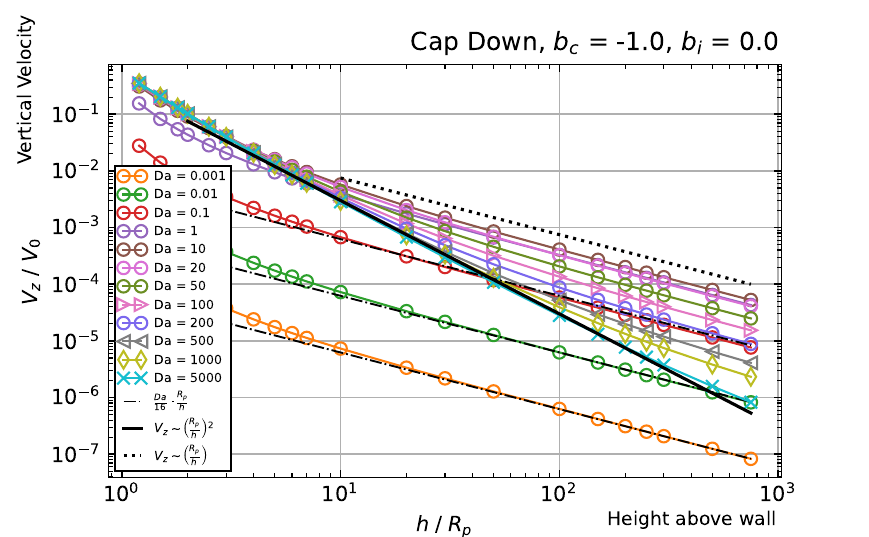}
    \caption{Dependence of the vertical velocity $V_z$ on the height $h$ for a cap-down sphere with non-uniform surface mobility near a circular patch with $R_d = R_p$.
    }
    \label{fig:non-uniform-cap-down}
\end{figure}
Now we consider a cap-down particle. For $b_i = -1$, $b_c = 0$, shown in Fig. \ref{fig:non-uniform-cap-down}, top panel, we obtain a similar behavior with that exhibited by a cap-up configuration with $b_i = 1$ and $b_c = 0$. For $b_i = 0$, $b_c = -1$, shown in in Fig. \ref{fig:non-uniform-cap-down}, bottom panel, we do not obtain any change in the sign of $V_z$. Accordingly, there are no hovering states for these parameters. However, for the high $\textrm{Da}$ curves, we still observe a transition from $V_z \sim \left(R_p/h\right)^2$ behavior at intermediate heights $h$, to  $V_z \sim \left(R_p/h\right)$ behavior at large heights $h$. The $V_z \sim \left(R_p/h\right)^2$ scaling is most evident for the largest value of $\textrm{Da}$, $\textrm{Da} = 5000$. This curve follows a $V_z \sim \left(R_p/h\right)^2$ scaling for a broad range of heights, with only a slight deviation for the largest height considered.  The transition to $V_z \sim \left(R_p/h\right)$ behavior is clearly evident for curves with $10 \leq \textrm{Da} \leq 200$.

\begin{figure}[!thb]
    \centering
\includegraphics[width=10cm]{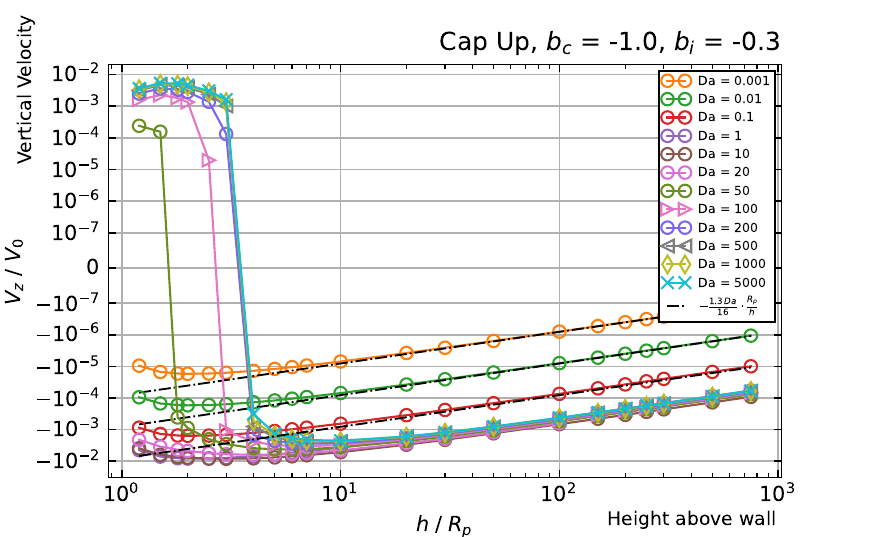}
    \caption{Dependence of the vertical velocity $V_z$ on the height $h$ for a cap-up sphere with $b_c = -1$ and $b_i = -0.3$ near a circular patch with $R_d = R_p$.}
    \label{fig:cap-up-bc-n1-bi-n03}
\end{figure}

Having characterized the cases $(b_i = \pm 1, b_c = 0)$ and $(b_i = 0, b_c = \pm 1)$, we now turn to scenarios in which both parameters are non-zero. As a specific example, Fig. \ref{fig:cap-up-bc-n1-bi-n03} shows $V_z$ vs $h$ for a cap-up sphere with $b_i = -0.3$ and $b_c = -1$. It can be seen that there is a crossover for all $\textrm{Da} \geq 50$. Moreover, the crossover heights are close to the wall: $h_c/R_p < 10$ for each curve with crossover. Thirdly, at crossover, the slope of $V_z$ vs. $h$ is negative, indicating that the hovering state is stable against vertical perturbations. Given these observations, we conclude that $b_i = -0.3$ and $b_c = -1$ are particularly favorable values of the surface mobility for realizing stable hovering states near the wall. Fig. \ref{fig:cap-up-bc-n1-bi-n03} also shows the scaling predicted by Eq. (\ref{eq:far_field_vz}) for $\textrm{Da} \ll 1$. 

\begin{figure}[!t]
    \centering    \includegraphics[width=1\linewidth]{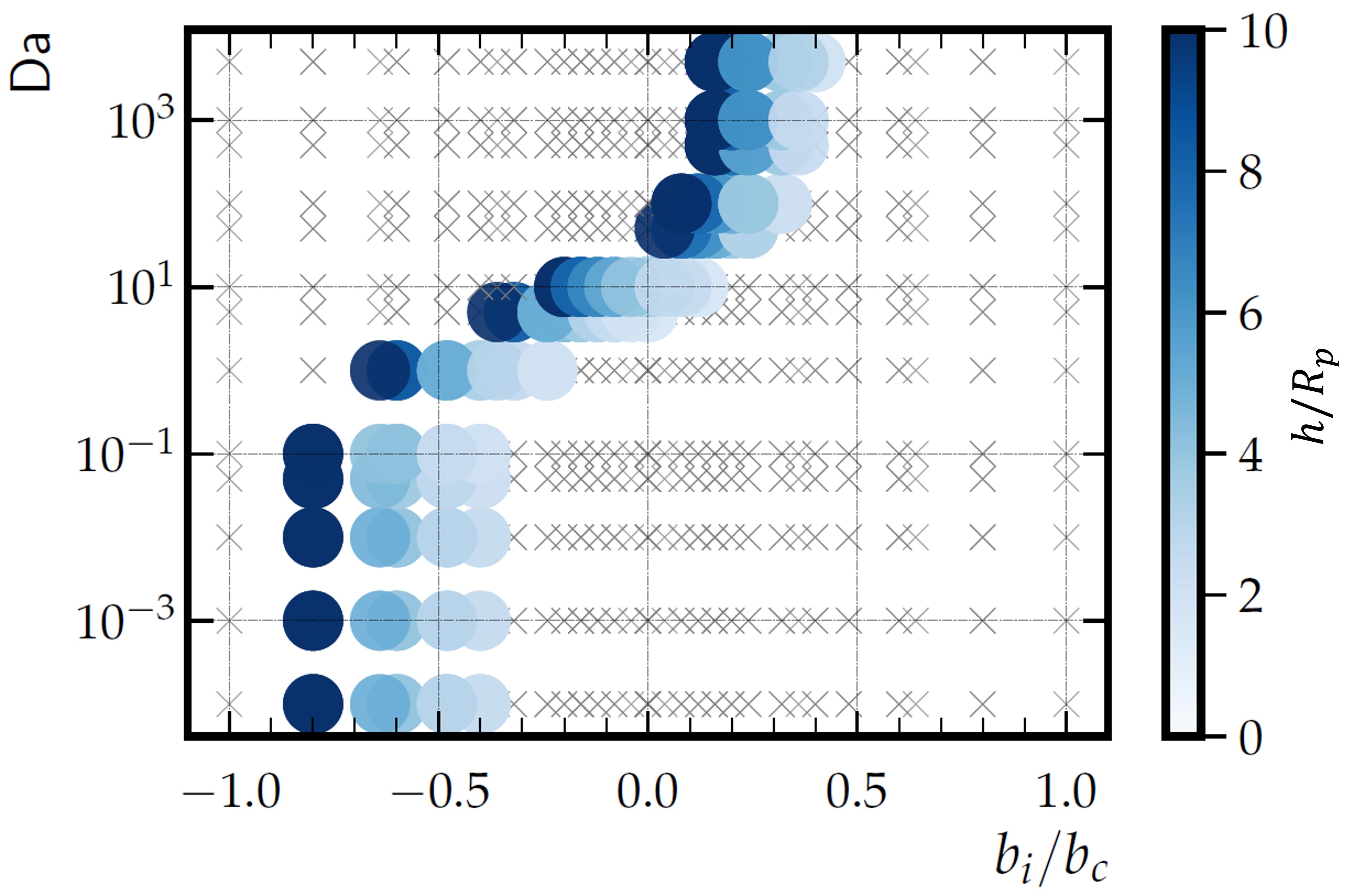}
    \caption{
    Hovering ``phase diagram'' illustrating the existence (symbols) and location (colour code) of hovering states as a function of the parameters $\textrm{Da}$ and $b_i/b_c$. The results correspond to a sphere in a cap-up orientation near a circular patch with $R_d = R_p$, and the phase diagram is restricted to hovering states with $1 \leq h/R_p \leq 10$.
    }
    \label{fig:phase_diagram}
\end{figure}

By varying $b_i$ and $b_c$ and identifying the existence and location of hovering states,  we determine a hovering ``phase diagram'' in the parameter space $b_i/b_c$ and $\textrm{Da}$, as shown in Fig. \ref{fig:phase_diagram} for a cap-up sphere near a circular patch with $R_d = R_p$. The ratio $b_i/b_c$ is chosen as the horizontal axis because the sign of $b_c$ only affects the vertical stability of the hovering state (i.e., the slope of $V_z$ vs. $h$ at the crossover point), and not the existence of a hovering state. Additionally, the phase diagram is restricted to hovering states that occur for $1 \leq h/R_p \leq 10$.  In comparison with the previously considered case of $b_i = 0$ and $b_c = 1$, for which the hovering state occurred for high values of $\textrm{Da}$, there are also hovering states for $\textrm{Da} < 1$ for $-1.0 < b_i/b_c < -0.3$.  Considering the product concentration field shown in Fig. \ref{fig:conc_product} for low $\textrm{Da}$ ($\textrm{Da} = 0.001$), it is apparent that these hovering states are not due to the ``inversion'' effect.  Instead, it is likely that they have the same origin as the hovering states demonstrated in Ref. \citenum{uspal2015self}, i.e., hydrodynamic and chemical interactions with the wall. Note that Ref. \citenum{uspal2015self} assumed zeroth-order chemical kinetics, which can be obtained from first-order kinetics in the limit $\textrm{Da} \rightarrow 0$. Additionally, Ref. \citenum{uspal2015self} considered an inert wall and uniform background concentration of ``fuel.''  In contrast with the present work, the problem considered in Ref. 
 \citenum{uspal2015self} gives hovering states in the range $1.02 \leq h/R_p \leq 7$ for $-1.4 \lesssim  b_i/b_c \lesssim -0.95$ for a Janus sphere that is half covered by catalyst. Therefore, although the $\textrm{Da} \ll 1$ hovering states obtained here may have the same physical origin as in Ref. \citenum{uspal2015self}, we find that sourcing the reactant field with a finite-sized patch can have a significant effect on the range of parameters that give a hovering state. Moreover, we note that the hovering states in Ref. \citenum{uspal2015self} are ``degenerate'' in that can occur at any position in the plane of the wall (i.e., any xy position), due to translational symmetry. In the present work, hovering states are localized directly above the patch ($x = 0$ and $y = 0$).

\begin{figure}[!th]
\includegraphics[width=1\linewidth]{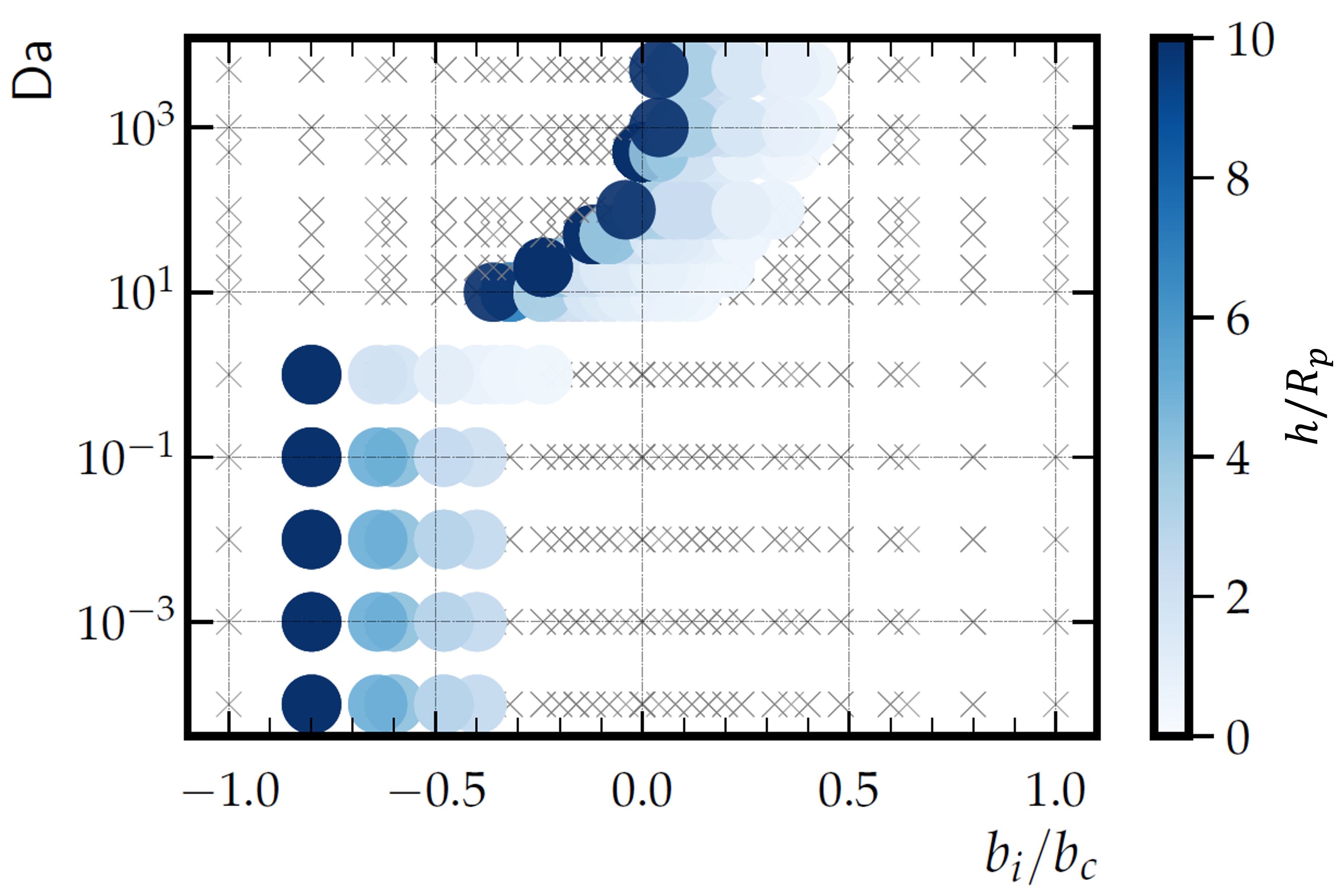}
\includegraphics[width=1\linewidth]{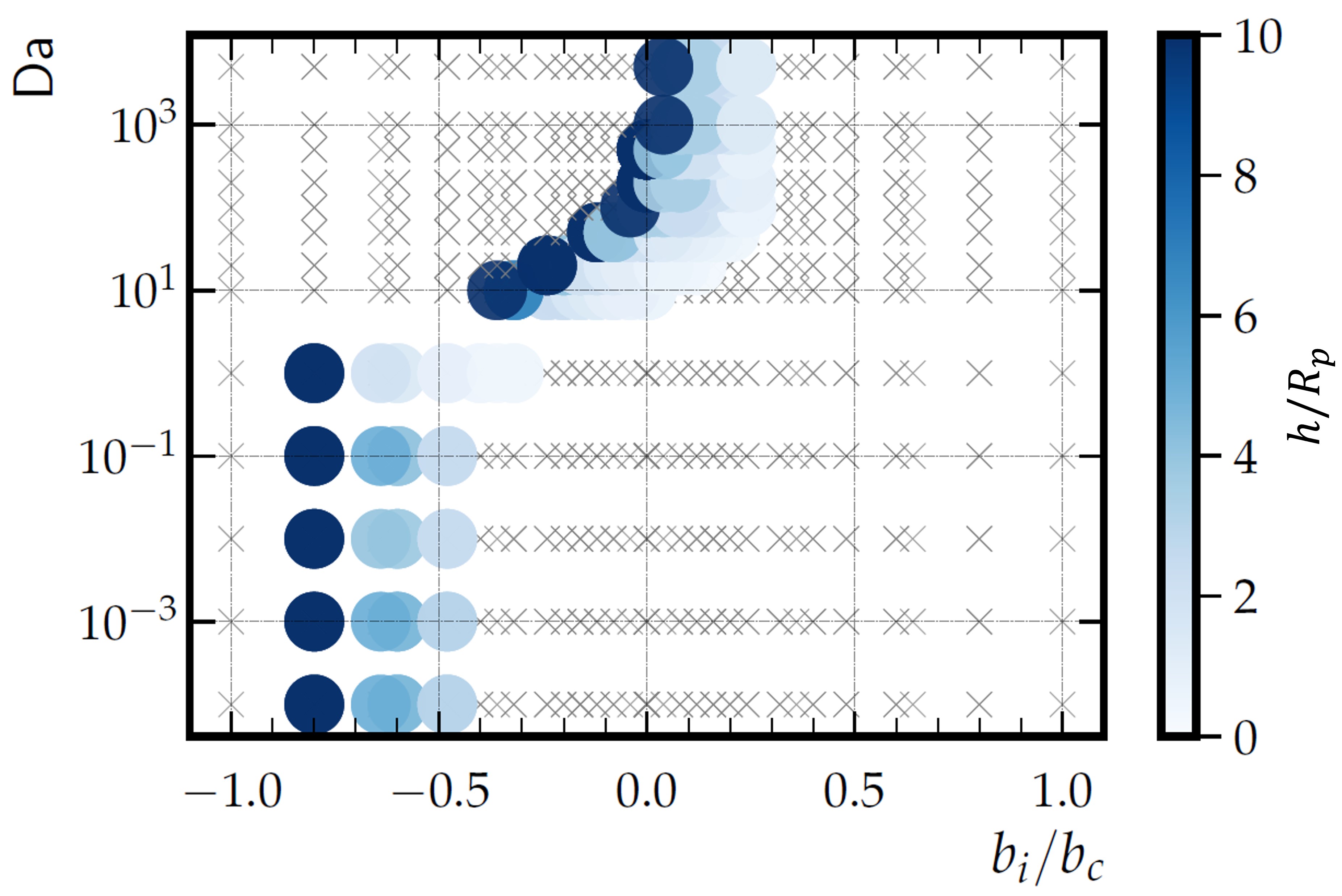}
    \caption{\label{fig:phase_diagram2} Hovering ``phase diagrams'' for a circular patch with radius $R_d = 1/3$ (above) and $R_d = 3$ (below). The diagrams illustrate  the existence (symbols) and location (colour code) of hovering states as a function of the parameters $\textrm{Da}$ and $b_i/b_c$. The phase diagrams are restricted to hovering states with $1 \leq h/R_p \leq 10$.
    }
    \label{fig:hovering_state_patch_radius}
\end{figure}

We also consider the phase diagrams for other patch sizes: $R_d/R_p = 1/3$ (Fig. \ref{fig:phase_diagram2}, top) and $R_d/R_p = 3$ (Fig. \ref{fig:phase_diagram2}, bottom). Notably, for the smaller patch, there is a larger range of $b_i/b_c$ that gives a hovering state, for both small and large $\textrm{Da}$. For the larger patch, the range of $b_i/b_c$ giving a hovering state at low $\textrm{Da}$ is more narrow than for $R_d/R_p = 1$. The range at high $\textrm{Da}$ is similar, although shifted to the left. 

\begin{figure}
    \centering
    \includegraphics[width=10cm]{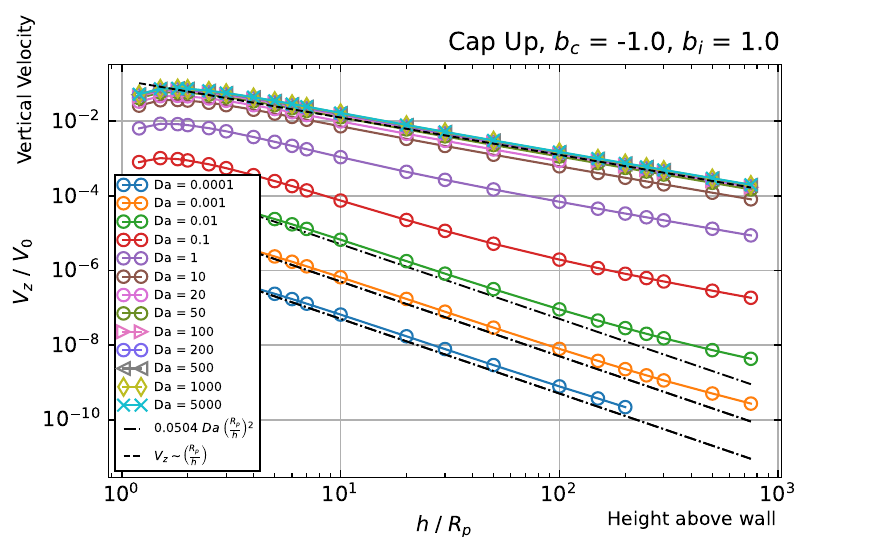}
    \caption{Dependence of the vertical velocity $V_z$ on the height $h$ for a cap-up sphere with $b_c = -1$ and $b_i = 1$ near a circular patch with $R_d = R_p$.}
    \label{fig:bc_n1_bi_1}
\end{figure}

The ``edge case'' $b_i = -b_c$ merits some additional consideration, as a particle with these mobility parameters in a uniform background concentration would be motionless in the limit $\textrm{Da} \rightarrow 0$. In other words, $\tilde{V}_{fs} = 0$. However, recalling that the patch creates a concentration gradient $\nabla c_r^b$, we expect a scaling law $V_z/V_0  \sim \textrm{Da} \, \left(\frac{R_p}{h}\right)^2$ for  $\textrm{Da} \rightarrow 0$.  We indeed recover this scaling  for $\textrm{Da} \ll 1$, as shown in  Fig. \ref{fig:bc_n1_bi_1}. (The numerical prefactor listed in the legend is determined from T{\u{a}}tulea-Codrean and Lauga \cite{tuatulea2018artificial}.) For high $\textrm{Da}$, there is no reason to expect these mobility parameters to give zero velocity in a uniform reactant field, and we indeed obtain a $V_z/V_0  \sim \left(\frac{R_p}{h}\right)$  scaling for $\textrm{Da} \gg 1$. For intermediate values of $\textrm{Da}$, there is crossover between the two scaling laws: $V_z/V_0  \sim \left(\frac{R_p}{h}\right)^2$ for small $h/R_p$, and  $V_z/V_0  \sim \left(\frac{R_p}{h}\right)$ for large $h/R_p$. This crossover is evident even for $\textrm{Da}$ as low as $\textrm{Da} = 0.001$. Given previous results on crossover, we expect this crossover to result from a competition between leading and subleading terms, although in this case for an expansion in powers of $\textrm{Da}$, instead of inverse powers of $\textrm{Da}$. We defer a more detailed examination to later work.

\section{Discussion}

Our analysis has focused on the dynamics for an axisymmetric configuration of the particle and the patch. A logical next step would be to investigate general three-dimensional motion. In particular, for the hovering states, an important question is whether they are stable against general three-dimensional perturbations, such as displacement lateral to the patch axis, or rotation of the particle axis away from the $\mathbf{\hat{z}}$ direction. 

\begin{figure}
    \centering
    \includegraphics[width=8cm]{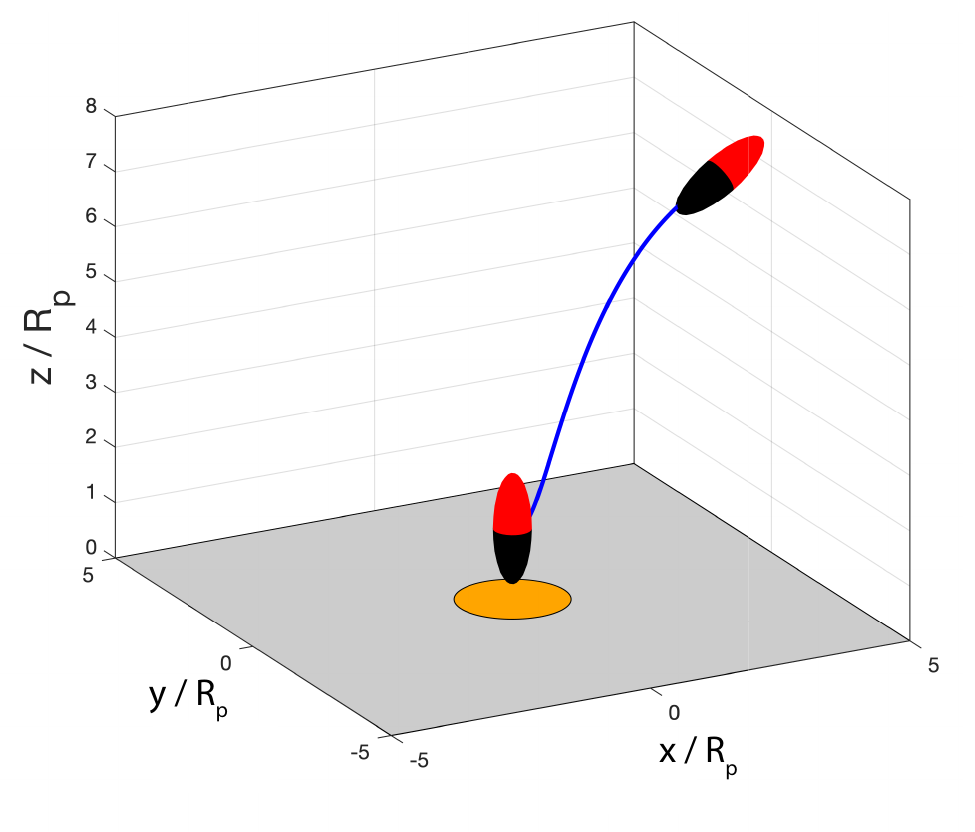}
    \includegraphics[width=7cm]{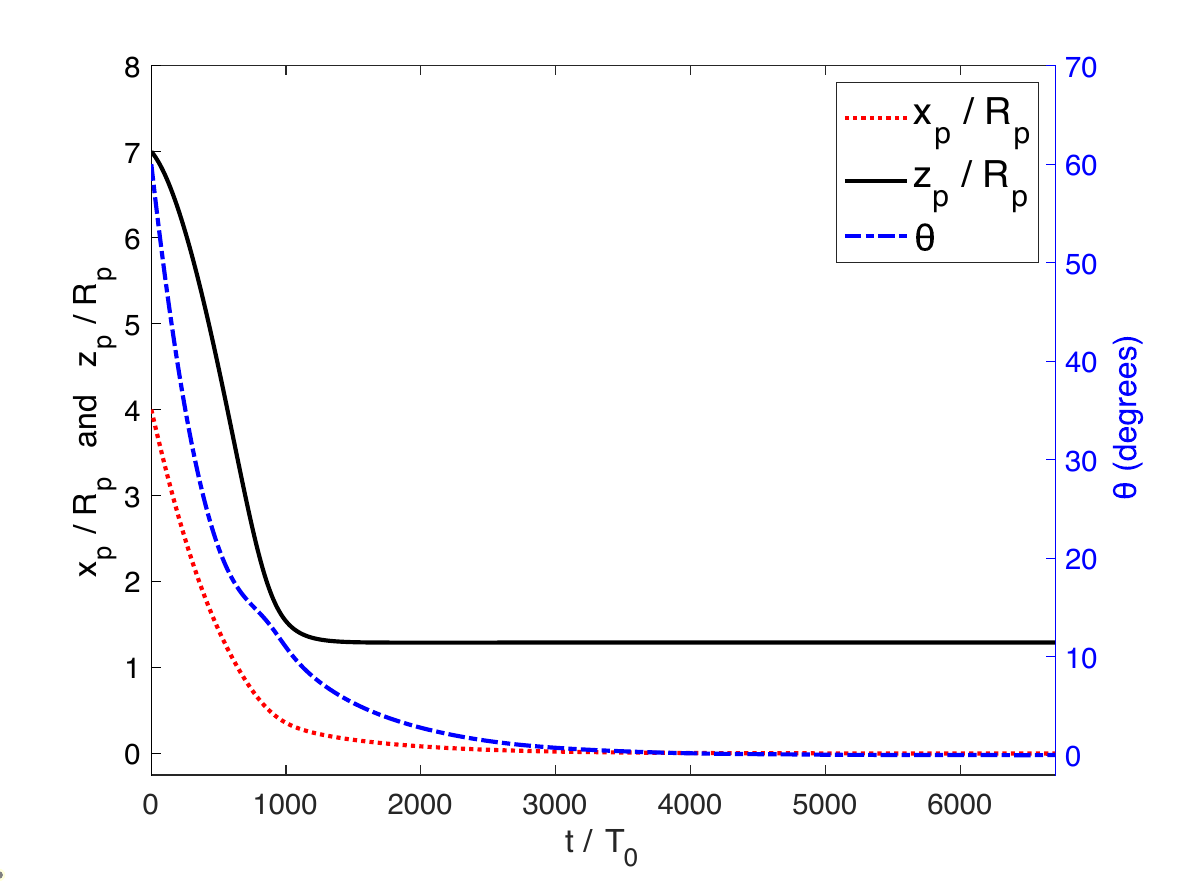}    \includegraphics[width=7cm]{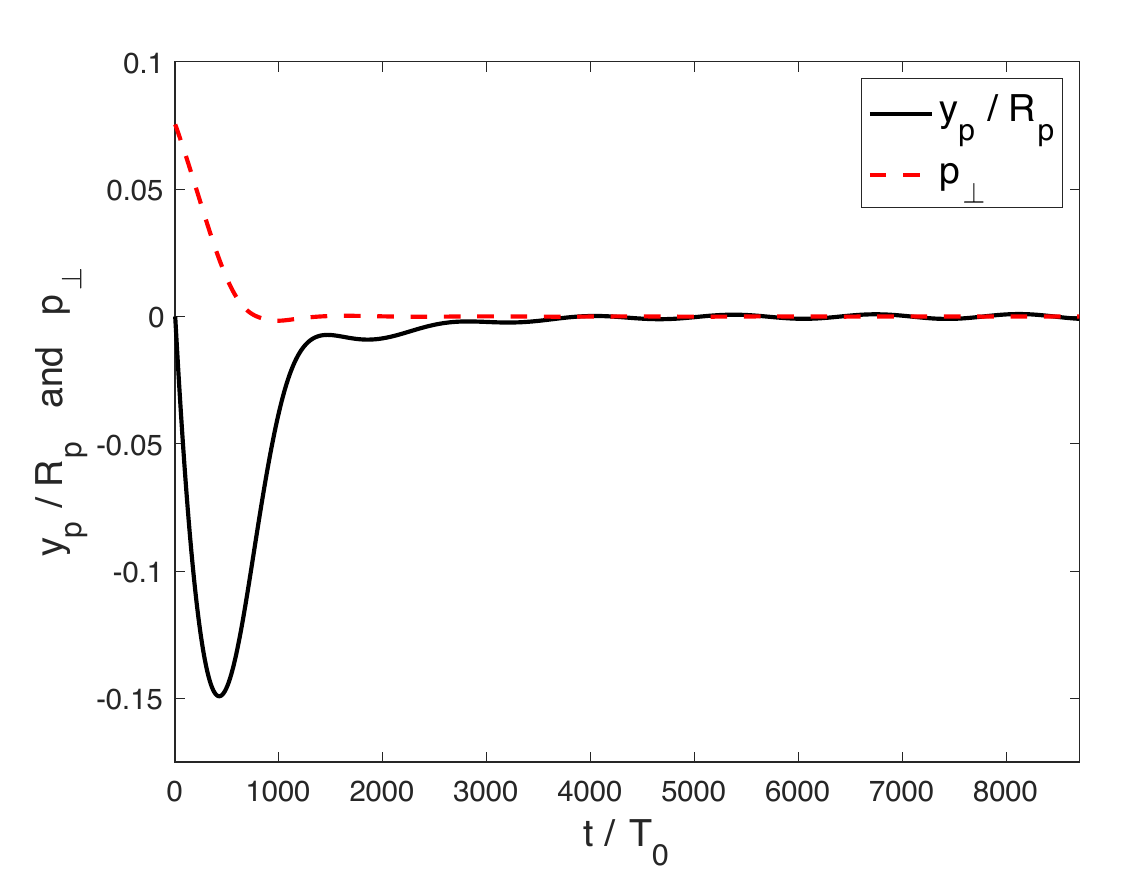}
    \caption{\label{rod_trajectory} (top panel) Three-dimensional trajectory of a prolate spheroid ($r_e = 1/3$, $b_i = -0.4$, $b_c = -1$, $\textrm{Da} = 100$) near a circular patch with $R_d = R_p$. The catalytic cap is shown in red. The particle is initially located off the patch axis ($x_p/R_p = 4$, $y_p/R_p = 0$, and $z_p/R_p = 7$.) Additionally, the particle is tilted towards the surface ($\theta = 60^{\circ}$) and slightly away from the patch ($\varphi = 5^{\circ}$), where $\theta$ and $\varphi$ are described in the text. The particle is attracted to a hovering state. (middle panel) Time evolution of the position components $x_p$ and $z_p$, as well as of the angle $\theta$.  (bottom panel) Time evolution of $y_p$ and $p_{\perp}$, the component of the particle orientation vector $\mathbf{\hat{p}}$ that is out of the plane defined by $\mathbf{\hat{z}}$ and the patch-to-particle vector $\mathbf{r}_p$.}
    \label{fig:rod_trajectory}
\end{figure}

As a first step towards this direction of investigation, we show in Fig. \ref{fig:rod_trajectory} an example of a three-dimensional trajectory, computed using a quaternion-based rigid-body dynamics method \cite{theers2016modeling}. This demonstrates that a prolate spheroid with phoretic mobilities $b_i = -0.4$ and $b_c = -1$, Damköhler number $\textrm{Da} = 100$, and aspect ratio $r_e = 1/3$ is attracted to the hovering state from an arbitrarily chosen initial position and orientation. The initial position of the particle ($x_p/R_p = 4$, $y_p/R_p = 0$, and $z_p/R_p = 7$) is located off the patch axis. Additionally, the initial particle orientation is neither aligned with the patch axis, nor with the plane defined by the patch axis and the patch-to-particle vector.  Specifically, the initial orientation of the spheroid is $\theta = 60^{\circ}$ and $\varphi = 5^{\circ}$. Here,  $\theta$ is the angle between the particle orientation vector $\mathbf{\hat{p}}$ and $\mathbf{\hat{z}}$, where $\mathbf{\hat{p}}$ is defined to point from the inert pole (black) to catalytic pole (red) of the particle. The quantity $\varphi$ is the angle between $\mathbf{\hat{p}}$ and $\mathbf{\hat{x}}$. Accordingly, all potential symmetries of the patch-and-particle configuration are broken. Despite this fact, the particle is able to steer to the patch, align its axis with the patch normal, and remain in the hovering state for the rest of the simulation. Fig. \ref{fig:rod_trajectory} shows the rod trajectory and the time evolution of its configuration. Since the azimuthal angle $\varphi$ is not well-defined when $\theta = 0$, we use $p_{\perp} = \mathbf{\hat{p}} \cdot (\mathbf{\hat{z}} \times \mathbf{\hat{r}}_p)$ to characterize the component of the orientation vector $\mathbf{\hat{p}}$ that is out of the plane defined by the patch normal $\mathbf{\hat{z}}$ and the particle position $\mathbf{r}_p$. Here, $\mathbf{\hat{r}}_p = \mathbf{r}_p/|\mathbf{r}_p|$.

\begin{figure}
    \centering
    \includegraphics[width=8cm]{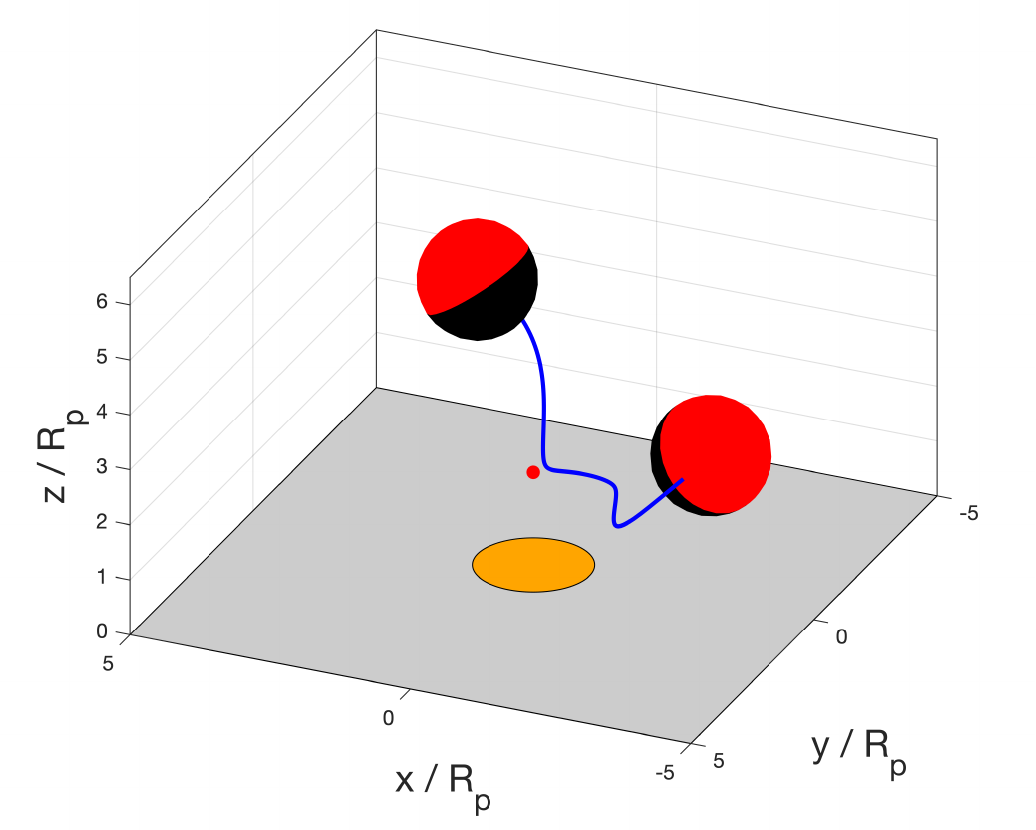}
    \caption{\label{fig:sphere_unstable} Three-dimensional trajectory of a sphere ($b_i = -0.2$, $b_c = -1$, $\textrm{Da} = 20$) near a circular patch with $R_d = R_p$. The red dot shows the spatial location of the hovering state. The sphere initially steers towards the hovering state, but slowly drifts away from the hovering state towards the edge of the patch. Once it reaches the vicinity of the edge, it quickly moves away from the patch. The initial conditions are $x_p/R_p = 1$, $y_p/R_p = 0$, $z_p/R_p = 5$, $\theta = 35^{\circ}$, and $\varphi = 5^{\circ}$. }
    \label{fig:enter-label}
\end{figure}

Interestingly, for spheres, we did not observe this long-time stability for all parameters tested. For some parameters, we found that the spheres could initially steer themselves towards the hovering state, but subsequently exhibited a slow drift away from the patch axis over a longer timescale. An example trajectory, illustrating such a behavior, is shown in Fig. \ref{fig:sphere_unstable}. This observation suggests that shape may have a crucial role in promoting stable ``docking,'' most likely by a tendency for elongated particles to align with nonlinear gradients \cite{solomentsev1995electrophoretic}. In future work, we will undertake a systematic examination of the three-dimensional linear stability of the hovering state as a function of  particle aspect ratio and surface mobilities. 

\section{Conclusions}

We have investigated the motion of a catalytic Janus particle activated by a patch that continuously releases reactant (``fuel.'') The patch has an axisymmetric shape and finite size, and is located on a solid planar wall that confines the liquid solution. The Janus particle has a spherical or spheroidal shape and an axisymmetric distribution of catalyst. Our model explicitly resolves transport of the fuel to the particle from the patch and its local consumption and depletion by the particle. We have shown that, under certain conditions of Damköhler number $\textrm{Da}$ (dimensionless rate of reaction), phoretic mobility (chemi-osmotic response of the particle), patch size, and other parameters (e.g., shape of the particle), there is a ``hovering'' state in which the particle remains motionless at a certain height above the patch. In the hovering state, the particle position is directly above the center of the patch, and the particle's axis of symmetry is aligned with the wall normal. Depending on the system parameters, this hovering state can be stable against perturbations of the particle position in the vertical direction. By using perturbation theory, we systematically examined the dependence of the hovering height on $\textrm{Da}$ for high Damköhler number. 

Our study focused on an axisymmetric configuration of the particle and patch. However, we laid the groundwork for consideration of general three-dimensional motion, and demonstrated that the hovering state can be stable against general three-dimensional perturbations. Specifically, we demonstrated that a particle with arbitrary initial orientation and position can steer to a chemical patch, i.e., exhibit chemotaxis. Three-dimensional motion would therefore be a natural direction for continuation of our work. This direction could address the phase diagram for three-dimensional stability of the hovering states, as well as how the system parameters can be tuned to enlarge the basin of stability, i.e., enlarge the range of initial orientations and positions from which a particle will be attracted to the hovering state. Continuation of our study could also probe the possible existence of other dynamical fixed points, such as continuous ``sliding'' along the edge of a circular patch. This sliding would be analogous to the orbiting observed for active colloids near geometric obstacles \cite{takagi2014hydrodynamic,brown2016swimming,van2023transition}. One can also imagine hovering states that are shifted away from the center of the patch. In this scenario, axisymmetry would imply that there is a continuous ring of degenerate hovering states. For a non-axisymmetric patch, such as square or star-shaped patch, there could be hovering states at a discrete set of spatial positions. Three-dimensional Brownian motion of the colloid can also be straightforwardly included in the framework of our model \cite{lisicki2014translational,katuri2018cross}.

Our results for a particle near a patch-decorated  planar wall may have implications for bound pair formation or ``docking'' of two freely suspended colloidal particles \cite{sharifi2016pair,nasouri2020exact,poehnl2023shape}. We recall that a planar wall can be regarded as a sphere with infinite radius of curvature. To our knowledge, a situation in which a reactant source is localized to one particle, and a catalytic region to a second particle, has not previously been considered in the literature, especially for first-order kinetics. Recalling that the hovering distance depends on Damköhler number in the particle/wall system, temporal variation of the rate of release in a particle/particle system could be exploited for temporally sequenced cargo pick-up and release \cite{palacci2013photoactivated}. Incorporation of additional chemical species and/or nonlinear chemical kinetics in our framework could allow for realization of more complex particle behaviors \cite{chen2024chemical}.

Finally, our findings may provide conceptual and technical guidance for advanced applications involving active colloids. In lab-on-a-chip systems, a chemical patch with a time-varying rate of release could be used for programmed ``trap-and-release'' of an active colloid. A planar surface decorated with many patches could template assembly of an active colloidal crystal. In materials science applications, a material that is impregnated with reactant, and which releases the reactant upon damage, could recruit active colloids to the damage site \cite{li2015self}. (Alternatively, the scenario can be that of locating corrosion spots on metal surfaces.) Additionally, for drug delivery applications, the chemical patch in this study could model an \textit{in vivo} biological source of reactant \cite{gao2014synthetic}. A stably hovering particle that continuously releases a drug would provide a highly concentrated and localized dose.  For biomedical applications, future consideration of aspects such as those of a non-Newtonian carrier fluid \cite{zhu2024self} or the effect of ambient flow \cite{katuri2018cross} would enhance the realism of our model.

\appendix

\section{
High Da Expansion}
We expand the (dimensionless) reactant field for $\textrm{Da} \gg 1$ as follows:
\begin{equation}
\tilde{c}_r(\tilde{\mathbf{r}}) = \tilde{c}_r^{(\infty)}(\tilde{\mathbf{r}}) + \frac{1}{\textrm{Da}} \tilde{c}_{r}^{(1)}(\tilde{\mathbf{r}}) +  \frac{1}{\textrm{Da}^2} \tilde{c}_{r}^{(2)}(\tilde{\mathbf{r}}) + \ldots
\end{equation}
The governing equation $\tilde{\nabla}^2 \tilde{c}_r = 0$ becomes
\begin{equation}
\tilde{\nabla}^2   \tilde{c}^{(\infty)}_r(\tilde{\mathbf{r}}) +  \frac{1}{\textrm{Da}} \tilde{\nabla}^2 \tilde{c}_{r}^{(1)}(\tilde{\mathbf{r}}) +  \frac{1}{\textrm{Da}^2} \tilde{\nabla}^2 \tilde{c}_{r}^{(2)}(\tilde{\mathbf{r}}) + \ldots = 0.
\end{equation} Therefore, each term $\tilde{c}_r^{(\ldots)}$ obeys Laplace's equation. The boundary condition on the catalytic cap becomes
\begin{equation}
\begin{split} 
\tilde{\nabla} \tilde{c}^{(\infty)}_r \cdot \mathbf{\hat{n}}  + \frac{1}{\textrm{Da}} \tilde{\nabla} \tilde{c}^{(1)}_r \cdot \mathbf{\hat{n}} + &  \frac{1}{\textrm{Da}^2} \tilde{\nabla} \tilde{c}^{(2)}_r \cdot \mathbf{\hat{n}} + \ldots = \\ &
\textrm{Da} \; \tilde{c}^{(\infty)}_r + \tilde{c}^{(1)}_r + \frac{1}{\textrm{Da}} \tilde{c}^{(2)}_r + \ldots
\label{eq:DaBC1}
\end{split}
\end{equation}
At $\mathcal{O}(\textrm{Da})$, we obtain $\tilde{c}^{(\infty)}_r = 0$ on the catalytic cap. We also note the boundary condition on the cap for the subleading problem, obtained at $\mathcal{O}(1)$ as $\tilde{c}_r^{(1)} = \tilde{\nabla} \tilde{c}^{(\infty)}_r \cdot \mathbf{\hat{n}}$.

On the inert side of the particle, we have 
\begin{equation}
\tilde{\nabla} \tilde{c}^{(\infty)}_r \cdot \mathbf{\hat{n}} + \frac{1}{\textrm{Da}} \tilde{\nabla} \tilde{c}^{(1)}_r \cdot \mathbf{\hat{n}} +  \frac{1}{\textrm{Da}^2} \tilde{\nabla} \tilde{c}^{(2)}_r \cdot \mathbf{\hat{n}}  +  \ldots = 0.
\end{equation}
Thus, for all fields $\tilde{c}_r^{(\ldots)}$, we obtain the no-flux boundary condition $\tilde{\nabla} \tilde{c}^{(\ldots)}_r \cdot \mathbf{\hat{n}} = 0$ on the inert face.

The boundary conditions for $\tilde{c}^{(\infty)}_r$ on the wall and at infinity are the same as given for $\tilde{c}_r$ in Section 3. Therefore, the inhomogeneity in the boundary condition on the wall for $\tilde{c}_r$ is accounted for in the leading order problem. For all $i \geq 1$, the boundary condition on the wall is $\tilde{\nabla} \tilde{c}^{(i)}_r \cdot \mathbf{\hat{n}}  = 0$, and the boundary condition at infinity is $\tilde{c}^{(i)}_r = 0$. 

Thus, we have a well-defined boundary value problem for $\tilde{c}^{(\infty)}_r$, with so-called mixed boundary conditions on the JP\cite{boymelgreen24} (i.e., Dirichlet on the cap and Neumann on the inert face.)

Now we turn to the product concentration. Again, we expand as follows:
\begin{equation}
\label{eq:cp_expansion}
\tilde{c}_p(\tilde{\mathbf{r}}) = \tilde{c}_p^{(\infty)}(\tilde{\mathbf{r}}) + \frac{1}{\textrm{Da}} \tilde{c}_{p}^{(1)}(\tilde{\mathbf{r}}) +  \frac{1}{\textrm{Da}^2} \tilde{c}_{p}^{(2)}(\tilde{\mathbf{r}}) + \ldots
\end{equation}
Again, each term $\tilde{c}_p^{(\ldots)}$ obeys Laplace's equation. The boundary condition on the catalytic cap becomes
\begin{equation}
\begin{split}
\tilde{\nabla} \tilde{c}^{(\infty)}_p \cdot \mathbf{\hat{n}} + \frac{1}{\textrm{Da}} \tilde{\nabla} \tilde{c}^{(1)}_p \cdot \mathbf{\hat{n}} + & \frac{1}{\textrm{Da}^2} \tilde{\nabla} \tilde{c}^{(2)}_p \cdot \mathbf{\hat{n}} + \ldots = \\ & -\textrm{Da} \; \tilde{c}^{(\infty)}_r - \tilde{c}^{(1)}_r - \frac{1}{\textrm{Da}} \tilde{c}^{(2)}_r +  \ldots
\end{split}
\end{equation}
At $\mathcal{O}(\textrm{Da})$, we recover the previous result that $\tilde{c}_r^{\infty} = 0$ on the catalytic face. At $\mathcal{O}(1)$, we have
$\tilde{\nabla} \tilde{c}_p^{(\infty)} \cdot \mathbf{\hat{n}} = -\tilde{c}_r^{(1)}$ on the cap. This boundary condition  can be simplified to remove dependence on the subleading term $\tilde{c}_r^{(1)}$. From Eq. (\ref{eq:DaBC1}), we obtain $\tilde{c}_r^{(1)} = \tilde{\nabla} \tilde{c}^{(\infty)}_r \cdot \mathbf{\hat{n}}$ on the cap. Therefore, the leading order term in the product flux on the cap can be obtained from the solution of the boundary value  problem for $\tilde{c}_r^{(\infty)}$ as $\tilde{\nabla} \tilde{c}_p^{(\infty)} \cdot \mathbf{\hat{n}} = -\tilde{\nabla} \tilde{c}^{(\infty)}_r \cdot \mathbf{\hat{n}}$. This result makes intuitive sense on physical grounds; consumption of reactant molecules entails  corresponding production of product molecules. Similarly, we can obtain $\tilde{\nabla} \tilde{c}_p^{(1)} \cdot \mathbf{\hat{n}} = -\tilde{\nabla} \tilde{c}^{(1)}_r \cdot \mathbf{\hat{n}}$ on the cap.

On the inert side of the particle, we have 
\begin{equation}
\tilde{\nabla} \tilde{c}^{(\infty)}_p \cdot \mathbf{\hat{n}} + \frac{1}{\textrm{Da}} \tilde{\nabla} \tilde{c}^{(1)}_p \cdot \mathbf{\hat{n}} + \frac{1}{\textrm{Da}^2} \tilde{\nabla} \tilde{c}^{(2)}_p \cdot \mathbf{\hat{n}} + \ldots = 0.
\end{equation} We again have $\tilde{\nabla} c^{(\ldots)}_p \cdot \mathbf{\hat{n}} = 0$ on the inert side for all fields $c^{(\ldots)}_p$. The boundary conditions on the wall and infinity for all fields $\tilde{c}_p^{(..)}$ are no-flux at the wall and vanishing concentration at infinity. 

Thus, we  have obtained a boundary value problem for $\tilde{c}_p^{(\infty)}$ with Neumann conditions on both sides of the JP, with  the prescribed product flux on the catalytic cap obtained from the solution to the problem for $\tilde{c}_r^{(\infty)}$.

The product concentration gradient determines the particle velocity. We define
\begin{equation}
V_z = V_z^{(\infty)} + V_z^{(1)} + V_z^{(2)} + \ldots    
\end{equation}
Here, for convenient comparison with numerical data, we choose to include the dependence on $\textrm{Da}$ in each term $V_z^{(..)}$. The dependence of each term $V_z^{(..)}$ on \textrm{Da} is given by the corresponding term in the series expansion in Eq. (\ref{eq:cp_expansion}).   Therefore, for example,  $V_z^{(\infty)}$ has no dependence on $\textrm{Da}$, $V_z^{(1)} \sim \textrm{Da}^{-1}$, etc.

At this point, we have established that, in the limit  $\textrm{Da} \rightarrow \infty$, the rate of production of product molecules on the cap $\mathbf{\hat{n}} \cdot \tilde{\nabla} \tilde{c}_p \sim \mathbf{\hat{n}} \cdot \tilde{\nabla} \tilde{c}_p^{(\infty)}$ no longer depends on $\textrm{Da}$. This result entails that, for a given set of mobility parameters, the $V_z$ vs. $h/R_p$ curves for different values of $\textrm{Da}$ should collapse onto a universal curve as $\textrm{Da} \rightarrow \infty$. This is observed for $\textrm{Da} \gg 1$ curves in various figures (e.g., Fig. \ref{fig:uniform-mobility-sphere}), although the values of $\textrm{Da}$ for which collapse is observed depends on the mobility parameters. Additionally, for some mobility parameters, collapse of a $\textrm{Da}$ curve may occur over a limited range of heights $h$. For instance, in Fig. \ref{fig:non-uniform-cap-down}, collapse is observed at lower values of $h/R_p$. The curves with higher $\textrm{Da}$ exhibit collapse for a larger range of $h/R_p$.

\subsection{Uniform phoretic mobility}

For the case of uniform phoretic mobility, we can obtain the scaling law for the universal curve, including the numerical prefactor, through the following arguments.
We approximate the JP as being in unconfined solution, immersed in a uniform background concentration $\tilde{c}^b_r(\tilde{\mathbf{r}}_p)$. We use the approximation in Eq. (\ref{eq:pointsource}) for the background reactant concentration. In dimensionless form, we have
\begin{equation}
\tilde{c}_r^b(\tilde{\mathbf{r}}_p) \approx \frac{\tilde{A}_d}{2} \frac{R_p}{h}.   
\end{equation}
Based on this approximation, we expect $V_z/V_0 \sim R_p/h$. Now we have obtained a mixed boundary value problem for $\tilde{c}_r^{(\infty)}$: $\tilde{\nabla}^2 \tilde{c}_r^{(\infty)} = 0$ in the solution, $ \tilde{c}_r^{(\infty)}(\tilde{\mathbf{r}}) \rightarrow \tilde{c}^b_r(\tilde{\mathbf{r}}_p)$ as $|\tilde{\mathbf{r}} - \tilde{\mathbf{r}}_p| \rightarrow \infty$,  $\mathbf{\hat{n}} \cdot 
 \tilde{\nabla} \tilde{c}_r^{(\infty)} = 0$ on the inert side of the JP, and $\tilde{c}_r^{(\infty)} = 0$ on the catalytic cap. This is the boundary value problem formulated by Davis and Yariv for a sphere in the $\textrm{Da} \rightarrow \infty$ limit in Ref. \citenum{davis2022self}. They obtain the coefficients $B_n$ in a harmonic expansion, $\tilde{c}_r^{(\infty)}(\tilde{\mathbf{r}}) = \tilde{c}^b_r(\tilde{\mathbf{r}}_p) + \sum_{n=0}^{\infty} B_n \tilde{r}^{-(n+1)} P_n(\cos \theta)$. Here, $\tilde{r} = |\tilde{\mathbf{r}} - \tilde{\mathbf{r}}_p|$ denotes distance from the JP center, non-dimensionalized by $R_p$; $P_n(x)$ is the Legendre polynomial of degree $n$; and $\theta$ denotes polar angle in a spherical coordinate system with its origin at the particle center. The polar angle $\theta$ is measured with respect to a vector pointing from the inert pole of the JP to the catalytic pole. 
 
 Davis and Yariv also obtain the Janus particle velocity for the case of a sphere with uniform phoretic mobility \cite{davis2022self}. However, they assume - in contrast with the present manuscript - that the slip velocity is driven by the surface gradient of the \textit{reactant} concentration, and \textit{not} by the gradient of a product concentration.  Since the product concentration field obeys Neumann conditions over the whole surface of the JP, we cannot immediately use this result. However, if we write the product concentration field as $\tilde{c}_p^{(\infty)} = \sum A_n \tilde{r}^{-(n+1)} P_n(\cos \theta)$, we obtain that $A_n$ = $-B_n$ for all $n$.  Therefore, their result that $V_z^{(\infty)}/V_0 = \pm 0.3 \, \tilde{c}_{r}^{b}(\tilde{\mathbf{r}}_p) $ applies here. (The sign of the velocity is determined by the cap-up or cap-down character of the particle.) We therefore obtain that $V_z/V_0 = \pm 0.15 \, \tilde{A}_d \, \frac{R_p}{h}$ as $\textrm{Da} \rightarrow \infty$. Aside from the sign, this result is independent of the cap-up or cap-down character of the particle, since it was obtained using the approximation of a uniform background concentration of reactant. 

\subsection{Crossover behavior for $b_c = 1$, $b_i = 0$}

Here, we consider the case $b_c = 1$ and $b_i = 0$, which gives a crossover between two scaling laws.

In order to obtain this crossover, we argue that, at leading order in $\textrm{Da}$ as $\textrm{Da} \rightarrow \infty$, the particle velocity is zero for a uniform background concentration of reactant. The reason is the following. We recall that in the leading order problem, i.e., the problem for $\tilde{c}_{r}^{(\infty)}$, the boundary condition on the catalytic cap is $\tilde{c}_{r}^{(\infty)} = 0$. Additionally, assuming a uniform background concentration of reactant $\tilde{c}_{r}^{b}(\tilde{\mathbf{r}}_p)$, and neglecting the confining wall, we recall that $\tilde{c}_{r}^{(\infty)}$ and $\tilde{c}_{p}^{(\infty)}$ can be written as harmonic expansions, with the coefficients related by $A_n = -B_n$. Since $\tilde{c}_{r}^{(\infty)} = 0$ on the cap, the coefficients $B_n$ are such as to precisely cancel  the background concentration $\tilde{c}_{r}^{b}(\tilde{\mathbf{r}}_p)$ on the cap.  In other words, on the catalytic cap, $\sum B_n \, P_n(\cos \theta) = -\tilde{c}_{r}^{b}(\tilde{\mathbf{r}}_p)$. From $A_n = -B_n$, it follows that, on the catalytic cap, $\tilde{c}_{p}^{(\infty)} = \sum A_n \, P_n(\cos \theta) = \tilde{c}_{r}^{b}(\tilde{\mathbf{r}}_p)$. 

At this point, we have obtained that the product concentration on the cap is uniform as $\textrm{Da} \rightarrow \infty$, for a uniform background concentration of reactant. In the case $b_c = 1$, $b_i = 0$, only the cap is osmotically responsive, i.e., we can only have slip on the cap. However, since the slip is proportional to the surface gradient of product concentration, we obtain that the slip is zero over the entire surface of the particle. Therefore, to leading order in $\textrm{Da}$ as $\textrm{Da} \rightarrow \infty$, the particle does not move in a uniform background concentration of reactant. 

However, we can still obtain motion in the leading order problem by considering the role of the second term in Eq. (\ref{eq:crb_expansion}), which represents a linear gradient in the background reactant concentration. At leading order, the reactant concentration can be written as $\tilde{c}_r^{(\infty)} =  \tilde{{\nabla}} \tilde{c}_{r}^b|_{\tilde{\mathbf{r}} = \tilde{\mathbf{r_p}}}  \cdot (\tilde{\mathbf{r}} - \tilde{\mathbf{r}}_p) + \sum_{n=0}^{\infty} C_n \tilde{r}^{-(n+1)} P_n(\cos \theta)$. We still have the boundary conditions $\tilde{c}_r^{(\infty)} = 0$ on the catalytic cap and $\mathbf{\hat{n}} \cdot 
 \tilde{\nabla} \tilde{c}_r^{(\infty)} = 0$ on the inert face.  Therefore, the values of $C_n$ must be such as to cancel the linearly varying background reactant concentration on the catalytic face. Concerning the product concentration, we can write this field as $\tilde{c}_p^{(\infty)} =  \sum_{n=0}^{\infty} D_n \tilde{r}^{-(n+1)} P_n(\cos \theta)$, with the same boundary conditions as before. By the same arguments, we obtain $D_n = -C_n$. Therefore, the product concentration varies with position on the catalytic face. This spatial variation is sufficient to give motion of the particle when $b_c = 1$ and $b_i = 0$. Since the background gradient $ \tilde{\nabla} \tilde{c}_{r}^b$ decays as $(R_p/h)^2$, we obtain that $V_z^{(\infty)}/V_0 \sim (R_p/h)^2$. The sign will depend on the cap-up or cap-down character of the particle, since the linear gradient in background reactant concentration is not isotropic. 

 We briefly argue that the subleading problem will give a contribution to velocity $V_z^{(1)}/V_0 \sim \textrm{Da}^{-1} (R_p/ h)$. Approximating the JP as being in unconfined solution, we have the boundary condition $\tilde{c}_r^{(1)}(\tilde{\mathbf{r}}) \rightarrow 0$ as $|\tilde{\mathbf{r}} - \tilde{\mathbf{r}}_p| \rightarrow \infty$, and we write $\tilde{c}_r^{(1)} = \sum E_n \, \tilde{r}^{-(n+1)} P_n(\cos \theta)$. We recall that the boundary conditions on the JP are $\tilde{c}_r^{(1)}  = \mathbf{\hat{n}} \cdot \tilde{\nabla} \tilde{c}_r^{(\infty)}$ on the catalytic cap, and $\mathbf{\hat{n}} \cdot \tilde{\nabla} \tilde{c}_r^{(1)} = 0$ on the inert face. This is a mixed boundary value problem, giving the dual series equations $\sum E_n \, P_n(\cos \theta) = -\sum (n+1) B_n \, P_n(\cos \theta)$ on the catalytic face, and $\sum E_n (n + 1) \, P_n(\cos \theta) = 0$ on the inert face, where the $B_n$ are known from solution of the leading order problem.
Now, for the particle to remain motionless, the reactant field must take a constant value on the catalytic face. We recall that the coefficients $B_n$ give a constant value of the reactant on the catalytic face and no-flux on the inert face. By uniqueness of solutions to linear equations, we deduce that $E_n = \alpha B_n$. (The constant $\alpha$ is included to allow for the scenario in which  $\tilde{c}_r^{(\infty)}$ and  $\tilde{c}_r^{(1)}$ take different uniform values on the cap.)
However, from the dual series equations, it is clear that $E_n$ cannot be a constant multiple $\alpha$ of $B_n$: $E_n \neq \alpha B_n$. Therefore, the field $\tilde{c}_r^{(1)}$ cannot take a uniform value on the cap and also satisfy the no-flux boundary condition on the inert face. Since the field $\tilde{c}_{p}^{(1)} = \sum F_n \tilde{r}^{-(n+1)} P_n(\cos \theta)$, where $F_n = -E_n$, it follows that $\tilde{c}_{p}^{(1)}$ is non-uniform on the cap. Therefore, in the subleading problem, the particle can move in a uniform background reactant concentration. We thus obtain $V_z^{(1)}/V_0 \sim \textrm{Da}^{-1} (R_p/ h)$.

 \section{
 \label{sec:App_bipolar} Solution in Bipolar Coordinates}

Since the set-up considered in this study possesses axial symmetry and involves Laplace or Stokes equations with boundary conditions on a sphere and on a wall, an analytic solution in terms of series in bipolar coordinates is possible as discussed in the followings.

\subsection{\label{bipolar_coord} Bipolar coordinates. Parametrization of the set-up}

The bipolar coordinates $(\xi,\eta)$ with $-\infty<\xi<\infty$ and $0\leq \eta \leq \pi$ are defined via the cylindrical coordinates $z \geq 0$ and $s \geq 0$ (the distance from the $z$-axis) as \cite{BrennerBook,Brenner1961}
\begin{equation}
  \label{eq:bipolar}
  z = \varkappa \frac{\sinh\xi}{\cosh\xi - \cos\eta} \qquad
  s = \varkappa \frac{\sin\eta}{\cosh\xi - \cos\eta},
\end{equation}
where 
\begin{equation}
\label{eq:def_kappa}
\varkappa =  R_p \sinh \xi_0
\end{equation}
and 
\begin{equation}
\label{eq:def_xi0}
\xi_0 = \mathrm{arccosh}(h/R_p)\,.
\end{equation}
These choices are such that the manifold $\xi = \xi_0$ corresponds to the spherical surface of radius $R_p$ centered at $z = h$ (i.e., the surface of the particle). The third coordinate is $\phi$ (azimuthal angle), defined in the usual way. Since the system in which one is interested has axial symmetry, we search for axisymmetric solutions for the disturbance of the density of reactant, $c_r^b$ and for the density of product molecules, $c_p$ as well as for the hydrodynamics. (Accordingly, there is no dependence on $\phi$, the possibility of symmetry-breaking solutions not being considered here).

The metric factors, which enter in the calculation of gradients and area elements (e.g., $\nabla = \be_\xi  g_\xi^{-1} \partial_\xi + \be_\eta  g_\eta^{-1} \partial_\eta + \be_\phi  g_\phi^{-1} \partial_\phi$), are given by
\begin{equation}
 \label{eq:metric_h}
 g_\xi = g_\eta = \frac{\varkappa}{\cosh \xi - \omega}\,,~~ g_\phi = g_\xi \sqrt{1-\omega^2} \,,~
 \mathrm{where~}\omega := \cos \eta\,.
\end{equation}

In terms of the bipolar coordinates system introduced above, the plane $z=0$ of the wall corresponds to $\xi = 0$, while the surface at infinity is parametrized by $\xi = 0, \eta =0~(\omega = 1)$.  The activity distribution $K(s) \mapsto K(\omega)$ is calculated by noting that $\xi = 0, \, s = 0$ (the center of the patch) corresponds to $\eta = \pi (\omega = -1)$ while $\xi = 0,\, s = R_d$ (the circumference of the patch) corresponds to 
\begin{equation}
\label{eq:def_om_d}
\omega_d  
= \frac{-(h/R_p)^2 + (R_d/R_p)^2 +1}{(h/R_p)^2 + (R_d/R_p)^2 - 1} 
\,.
\end{equation}
Accordingly, 
\begin{equation}
\label{eq:def_K_om}
K(\omega) = \begin{cases} 
1, \quad -1 \leq \omega \leq \omega_d\\
0, \quad \omega_d < \omega \leq 1
\end{cases}.
\end{equation}

Similarly, the activity distribution $f(\br_J) \mapsto f(\omega)$, where $\br_J$ denotes positions on the surface of the Janus particle, is calculated by noting that: (i) the ``South'' pole $(s = 0, z = h - R)$ corresponds to $(\xi = \xi_0, \omega = -1)$, (ii) the ``North'' pole $(s = 0, z = h + R)$ corresponds to $\xi = \xi_0, \omega = +1$; and (iii) the ``Equator'' $\{s = R, z = h\}$ corresponds to $\xi = \xi_0$, 
\begin{equation}
\label{eq:def_omegaE}
\omega_e = (\cosh\xi_0)^{-1} = R_p/h.
\end{equation} 
Accordingly, the distribution $f(\br_J) \to f^{(cu),(cd)}(\omega;\xi_0)$ (corresponding to the "cap up" and "cap down" configurations; only here we indicate explicitly, as a reminder, that this distribution depends on the location of the sphere) is given by
\begin{eqnarray}
\label{eq:def_f_up_down}
&&f^{(cu)}(\omega) = \begin{cases} 
0, -1 \leq \omega \leq \omega_e \\
1, \omega_e < \omega \leq 1
\end{cases}, \nonumber\\
&&f^{(cd)}(\omega) = \begin{cases} 
1, -1 \leq \omega \leq \omega_e \\
0, \omega_e < \omega \leq 1
\end{cases}.
\end{eqnarray}
Finally, the normal unit vectors (oriented into the fluid) at the wall and at the particle, respectively, $\bn_w$ and $\bn_p$, are given by $\bn_w = \be_z = \be_\xi$ and $\bn_p = -\be_\xi$.

\subsection{\label{sec:diff_react} The disturbance concentration of reactant, $c_r^d(\br)$, in bipolar coordinates}

The disturbance field $c_r^d$ is the solution of Laplace equation subject to a vanishing field BC at infinity, the Robin BC at the surface of the particle expressed in \eq{chem3}, and the zero normal current BC at the wall (the activity at the patch being accounted for by the BC on $c_r^b$). In bipolar coordinates, the solution of the Laplace equation obeying the vanishing BC at infinity can be expressed in terms of the Legendre polynomials $P_n$ as \cite{Jeffery1912}
\begin{eqnarray}
  \label{eq:cr_bipolar}
  &&c_r^{d}(\br) = 
  C_0 (\cosh\xi - \omega)^{1/2} 
\times \\
&&
\sum_{n \geq 0} {  \left\lbrace 
    A_n \sinh \left[\left(n+\frac{1}{2}\right)\xi \right] + B_n 
\cosh \left[\left(n+\frac{1}{2}\right) \xi \right] \right \rbrace 
}
P_n(\omega)
\,, \nonumber 
\end{eqnarray}
The dimensionless coefficients $A_n$ and $B_n$ are determined from the two remaining boundary conditions as follows. By noting the orthogonality of the Legendre polynomials,
\begin{equation}
\label{eq:Leg_ortho}
\int\limits_{-1}^{1} d\omega P_n(\omega)P_m(\omega) = \frac{2}{2n+1}\delta_{n,m}\,,
\end{equation}
the zero normal current boundary condition at the wall ($\xi = 0$) straightforwardly implies 
\begin{equation}
\label{eq:sol_An}
A_n = 0 \,, n \geq 0\,.
\end{equation}
By using this result and the orthogonality of the Legendre polynomials, the boundary condition at the surface of the JP renders for the auxiliary unknowns 
\begin{equation}
x_n:=B_n \cosh \left[\left(n+\frac{1}{2}\right) \xi_0 \right],
\end{equation}
the following infinite system of linear equations:  
\begin{equation}
\sum\limits_{n \geq 0} H_{kn} x_n = \int\limits_{-1}^{1} d\omega \,T(\omega;\xi_0,{\cal R},Da) P_n(\omega)\,,~ k = 0,1,\dots 
\end{equation}
where
\begin{eqnarray}
\label{eq:def_hkn}
&&H_{kn}:=\int\limits_{-1}^{1} d\omega \,P_n(\omega)\,P_k(\omega)\,
\left\lbrace \frac{\sinh \xi_0}{2} \left(\cosh \xi_0 -\omega \right)^{1/2} \right. \\
&&+\left. \mathrm{Da} \left(n+\frac{1}{2}\right) f(\omega)\left(\cosh \xi_0 -\omega \right)^{3/2} \tanh \left[\left(n+\frac{1}{2}\right) \xi_0 \right] 
\right\rbrace\,,\nonumber
\end{eqnarray}
\begin{equation}
\label{eq:def_T}
T(\omega;\dots):= - \sinh \xi_0 
\left[\frac{R_d}{R_p} \,\be_\xi \cdot \nabla {\tilde c}_r^{b}(\br_J)+ Da \,f(\omega)\, {\tilde c}_r^{b}(\br_J)
\right]\,,
\end{equation}
where $\br_p=(\xi_0,\omega)$. In practice, the system is solved by truncating at a sufficient large order $N_{max}$, i.e., setting $B_{n>N_{max}} = 0$. \newline
For the results presented in the manuscript we have used $N_{max} = 10$ and numerically solved the system using the software Mathematica (version 14.0); this was sufficient to pass the cross-check  against the corresponding BEM results (see the good agreement between the two shown by the results in \fig{fig:conc_reactant}. If very accurate calculations of the concentrations are needed, additional analytical manipulations of the equations derived above may be necessary, as discussed by Ref. \citenum{Boymelgreen2024}.

\subsection{\label{sec:diff_prod} Concentration of product in bipolar coordinates}

Similarly with the case of the boundary value problem for the disturbance density field, the solution of the Laplace equation for the density of product, accounting for the BC at infinity (vanishing concentration) and for the BC at the wall (zero normal current), is expressed as
\begin{equation}
  \label{eq:cp_bipolar}
  c_p(\br) = 
  C_0 (\cosh\xi - \omega)^{1/2} 
\times \sum_{n \geq 0}
    F_n 
\cosh \left[\left(n+\frac{1}{2}\right) \xi \right] 
P_n(\omega)\,,
\end{equation}
The boundary condition at the surface of the JP (\eq{e10}) takes the form 
\begin{equation}
D_p \left[\frac{1}{g_\xi} \frac{\partial c_p(\xi,\omega)}{\partial \xi}\right]_{\xi = \xi_0} = + F c_r(\xi_0,\omega) f(\omega)\,;\nonumber
\end{equation}
With ${\tilde c}_r(\br): = c_r(\br)/C_0$ and 
\begin{equation}
\label{eq:def_Dratio}
{\cal D} := D_r/D_p\,,
\end{equation}
after employing the recursion relation \cite{GrRy94}
\begin{equation}
 \left(n + \frac{1}{2}\right)  \omega P_n(\omega) = \frac{(n+1)P_{n+1}(\omega) + n P_{n-1}(\omega)}{2}\,,
\end{equation}
and a couple of straightforward algebra manipulations to convert products of hyperbolic functions into sums of hyperbolic functions, the lhs can be re-arranged in a single series in Legendre polynomials. By projecting on the corresponding Legendre polynomial (using the orthogonality relation) it renders
\begin{eqnarray}
 \label{eq:sol_Fn}
 && (n+1) (F_n - F_{n+1}) \sinh\left[\left(n + \frac{3}{2}\right)\xi_0\right] \nonumber\\
 && + 
  n (F_n - F_{n-1}) \sinh\left[\left(n - \frac{1}{2}\right)\xi_0\right] 
  \\
&& = \underbrace{\left(2 n + 1\right)\, {\cal D}\,  \mathrm{Da}\, \int\limits_{-1}^{1} 
 \frac{f(\omega) {\tilde c}_r(\xi_0,\omega) P_n(\omega)}{(\cosh \xi_0 -\omega)^{1/2}}}_{:=p_n}\,,~~n \geq 0 \,,\nonumber
 \end{eqnarray}
with the convention $F_{-1} = 0$. Recalling that ${\tilde c}_r(\xi_0,\omega;\mathrm{Da})$ is a known function (calculated in the previous section), the quantity $p_n$ in the rhs is known (as also noted in the previous Subsections, additional analytical simplifications may be possible, see Ref. \citenum{boymelgreen24}) and \eq{eq:sol_Fn} turns into an infinite system of linear equations.  The system is solved by truncating at sufficiently large $n$, i.e., by setting $F_{n>N_{max}} = 0$ and by employing the auxiliary variable $r_n = F_n - F_{n-1}$: at $n = 0$ one directly obtains $r_1$, and then the $r_n$ is solved recurrently. The $F_n$s are then obtained by first summing up all the $r_{n>0}$, which gives $F_0$, and subsequently $r_n$ by recurrence from their definition and the known $r_{n}$. This concludes the derivation of the density $c_p(\br)$ of product molecules. As for the disturbance of the reactant concentration, the results are validated by the cross-check against the corresponding BEM results, which shows a very good agreement between the two methods (see \fig{fig:conc_product}).

\subsection{\label{sec:vel_bipolar} Velocity of the particle in terms of bipolar coordinates}

The calculation of the velocity of the particle, which has a prescribed 
phoretic slip actuation $\bv_s(\br)$ on its surface, near a no-slip wall follows from the expression, \eq{F4}, derived via the Lorentz reciprocal theorem. Since this calculation has been discussed in detail by Refs. \citenum{Dominguez2016,Malgaretti2018}, here we provide only a succinct description.

Owing to the symmetry of the system, only motion along the z-axis occurs; accordingly, only one auxiliary problem is needed in \eq{F4}, and this is chosen as the motion of a chemically inert spherical particle, of same radius $R_p$ and located at the same height $h$ with velocity $\bV' = V' \be_z$. This was studied by Ref. \citenum{Brenner1961}, which provides the force $\bF_a$ in the form of the well-known wall-correction of the Stokes formula and outlines the solution for the hydrodynamic flow in terms of a stream function $\Psi_a(\xi,\omega:=\cos \eta) = V' R_p^2 \psi_a(\xi,\omega)$, where $0 \leq \xi \leq \xi_0$ while $\psi_a(\xi,\omega)$ is dimensionless, which admits the series representation in bipolar coordinates as \cite{Jeffery1912,Brenner1961}:
\begin{eqnarray}
\label{eq:stream_func}
 && (\cosh\xi-\omega)^{3/2} \psi(\bs) = \nonumber\\ 
 &&\sum_{n=1}^{+\infty}
   \left\lbrace
  K_n \, {\cosh \left[\left(n-\frac{1}{2}\right)\xi\right]}
    + L_n \, {\sinh \left[\left(n-\frac{1}{2}\right)\xi\right]}
  \right. 
\nonumber \\
 && \left. +  M_n \, {\cosh \left[\left(n+\frac{3}{2}\right)\xi\right]}
    + N_n \, {\sinh \left[\left(n+\frac{3}{2}\right)\xi\right]}
  \right\rbrace 
\times  {\cal G}_{n+1}^{-1/2}(\omega) \,,
\end{eqnarray}
where
\begin{equation}
 {\cal G}_{n}^{-1/2}(\omega) = \dfrac{P_{n-2}(\omega) - P_{n}(\omega)}{2 n - 1}
\end{equation}
denotes the Gegenbauer polynomial of order $n$ and degree $-1/2$ {\cite{Brenner1961}}. The dimensionless coefficients $K_n$, $L_n$, $M_n$, and $N_n$ depend on $\xi_0$ and are determined by the boundary conditions at the wall and at the particle for the velocity field. For the no-slip boundary conditions corresponding to the auxiliary problem considered here, these coefficients are determined analytically in closed form \cite{Brenner1961}; the expressions, which are somewhat cumbersome and not particularly insightful, can be found in, e.g.,  Refs. \citenum{Brenner1961,Malgaretti2018}, and thus we do not list them here.

The phoretic slip velocity in the rhs of \eq{F4} is given by
 \begin{eqnarray}
  \label{eq:phor_slip}
 \bv_s(\br_J) &=& - b_0 b(\br_J) \nabla_{||} c_p(\br_J) \\
 &=& - V_0 b(\br_J) \left.\left(\frac{R_p\sqrt{1-\omega^2}}{g_\eta} \frac{\partial {\tilde c}_p}{\partial \omega}\right)\right|_{\xi = \xi_0} \be_\eta \,;
\end{eqnarray}
the derivative on the rhs is calculated analytically, using the series representation derived in the previous Subsection.

Since the unit vector along the direction normal to the surface of the particle is $\bn = -\be_\xi$, the integrand in the rhs of \eq{F4} will involve only the contraction $
\be_\xi \cdot {\bm{\sigma}}_a \cdot \be_\eta$ of the shear stress tensor of the auxiliary problem with the unit vectors $\be_{\xi,\eta}$, evaluated at $\xi = \xi_0$. This contraction is given by (see also Ref. \citenum{Malgaretti2018}),
\begin{equation}
\label{eq:stress_contr}
 \be_\xi \cdot {\bm{\sigma}}_a \cdot \be_\eta = \frac{\mu}{g_\xi}
 \left[\left(\frac{\partial u'_\xi}{\partial \eta} + \frac{\partial u'_\eta}{\partial \xi}\right) - \frac{1}{g_\xi} \left(u'_\eta \frac{\partial g_\eta}{\partial \xi} + u'_\xi \frac{\partial g_\xi}{\partial \eta}\right) \right]
\end{equation}
The velocity components $u'_{\xi,\eta}$ of the auxiliary problem, which are needed for the calculation of the contraction, are in turn evaluated from the stream function of the auxiliary problem as (see Ch. 4-4 in Ref. \citenum{BrennerBook})
\begin{subequations}
\label{eq:u_xi_eta}
\begin{equation}
u'_\xi = - \dfrac{1}{s\, g_\eta} \dfrac{\partial \Psi_a}{\partial \eta} 
\,,
\end{equation}
and
\begin{equation}
u'_\eta = \dfrac{1}{s\, g_\xi} \dfrac{\partial \Psi_a}{\partial \xi} 
\,.
\end{equation}
\end{subequations}

After performing the derivatives analytically, the integral in the rhs of \eq{F4} is computed numerically and the integrand is approximated by truncating the series representations corresponding to the phoretic slip and of the stress-tensor contraction to the first $N_{max} = 30$ terms, which seems sufficient as long as $h/R_p \gtrsim 1.1$. 

\begin{figure}
    \centering
    \includegraphics[trim={2cm 1cm 2cm 1.5cm},clip,width=8cm]{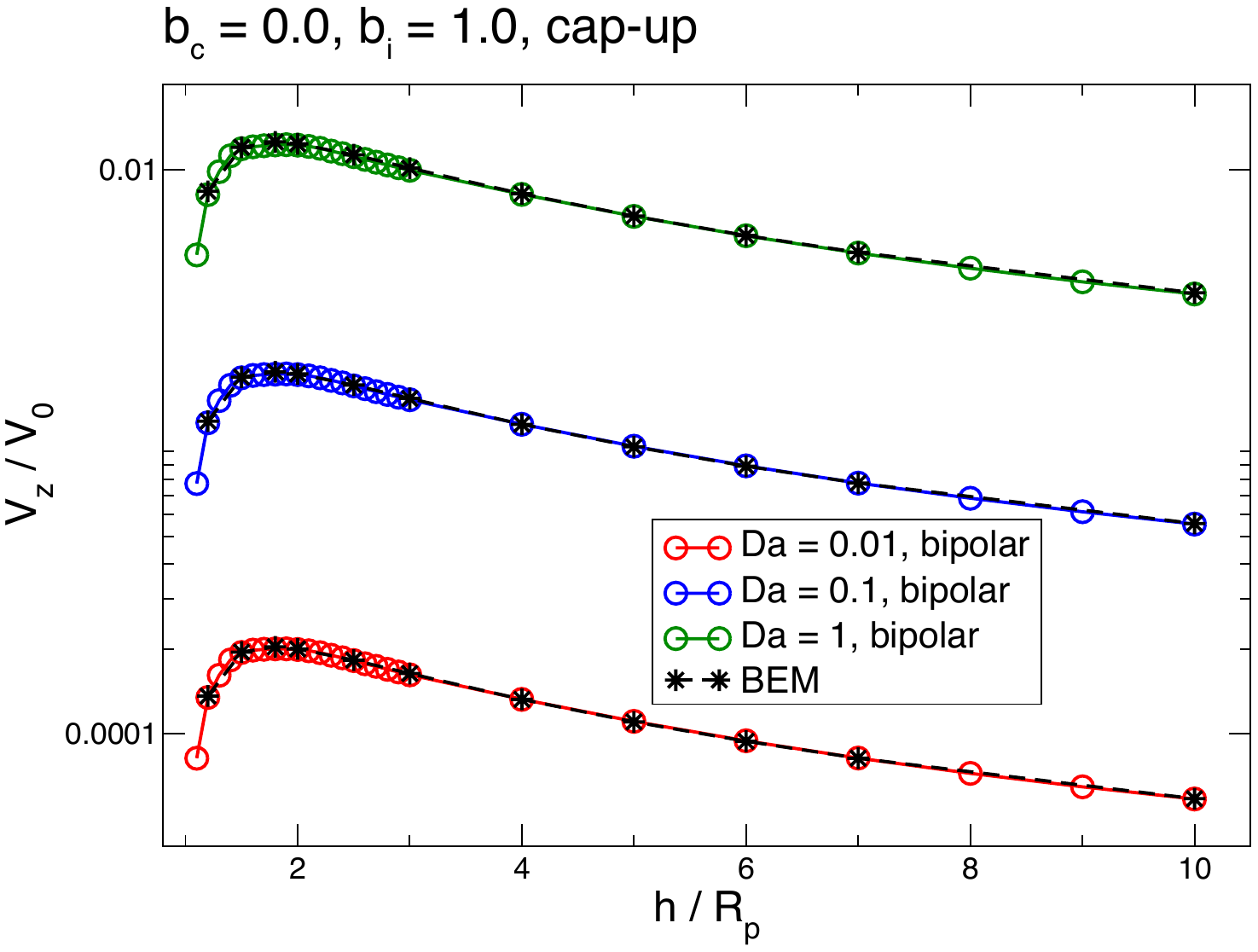}
        \includegraphics[trim={2cm 1cm 2cm 1.5cm},clip,width=8cm]{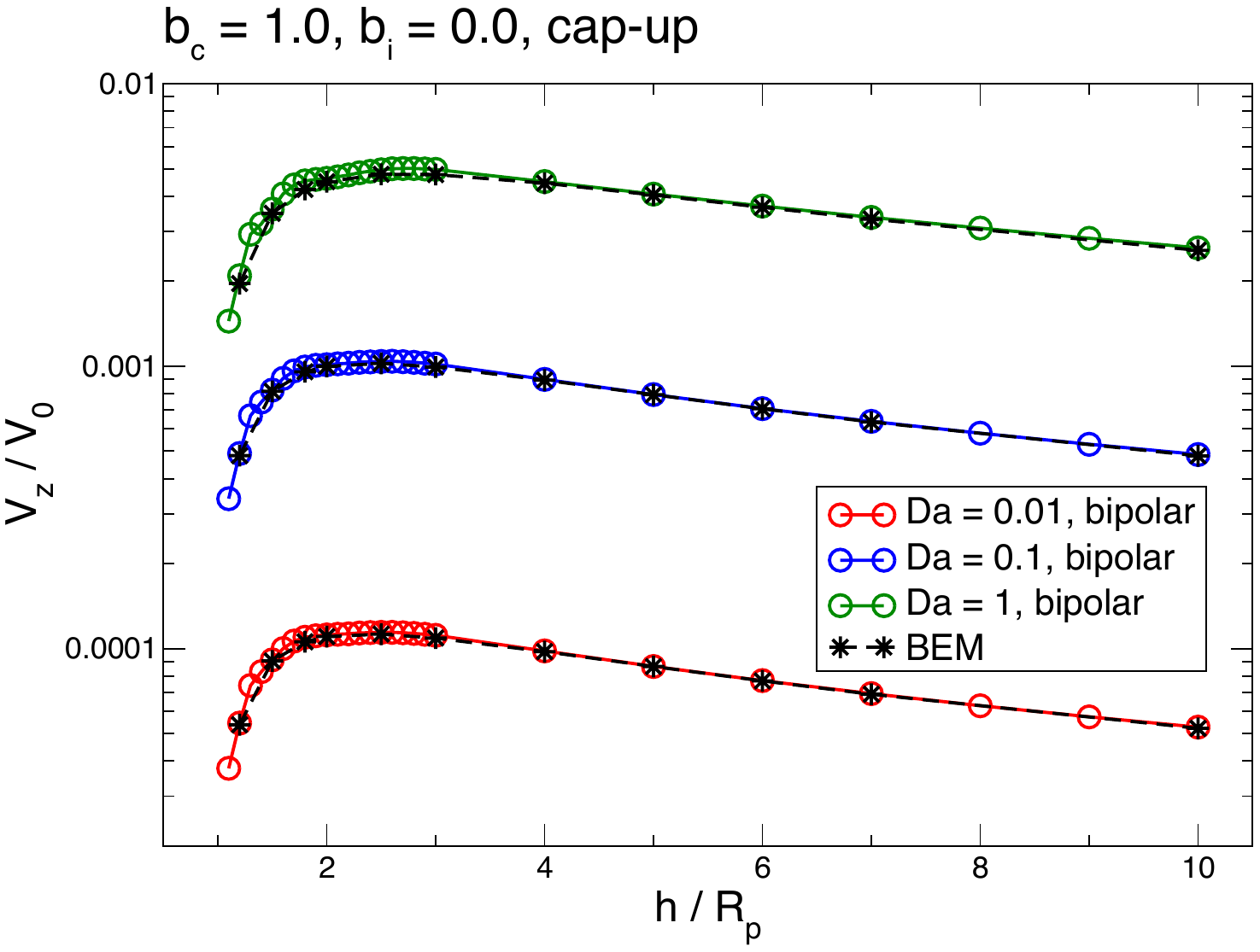}
    \caption{Velocity of a cap-up sphere with non-uniform surface mobility near a circular patch with radius $R_d = R_p$. The velocity is computed with the boundary element method and using bipolar coordinates.}
    \label{fig:bipolar_BEM_velocity}
\end{figure}

In Fig. \ref{fig:bipolar_BEM_velocity}, we show the velocity of a cap-up sphere with non-uniform surface mobility near a circular patch with radius $R_d = R_p$, computed with the boundary element method and using bipolar coordinates. The two methods show close quantitative agreement for the range of $h/R_p$ and values of $\textrm{Da}$ shown. The system defined by truncating Eq. (\ref{eq:sol_Fn}) to $n$ equations is observed to have slow convergence with $n$ for larger $h/R_p$ and higher $\textrm{Da}$. Accordingly, we use the BEM for velocity data shown in Section III.

\section*{Acknowledgments}
The technical support and advanced computing resources from University of Hawaii Information Technology Services – Cyberinfrastructure, funded in part by the National Science Foundation CC* awards \#2201428 and \#2232862 are gratefully acknowledged. W. E. U. and V. M.  gratefully acknowledge the Donors of the American Chemical Society Petroleum Research Fund for support of this research, grant number 60809-DNI9. W. E. U. and V. M. also gratefully acknowledge support from the Army Research Office under Grant Number W911NF-23-1-0190. 
M.N.P. acknowledges financial support through grants ProyExcel\_00505 funded by Junta de Andaluc{\'i}a, PID2021-126348NB-I00 funded by MCIN/AEI/10.13039/501100011033 and ``ERDF A way of making Europe'', and a Mar{\'i}a Zambrano grant from Ministerio de Universidades. The views and conclusions contained in this document are those of the authors and should not be interpreted as representing the official policies, either expressed or implied, of the Army Research Office or the U.S. Government. The U.S. Government is authorized to reproduce and distribute reprints for Government purposes notwithstanding any copyright notation herein.

\bibliography{patch} %

\end{document}